\documentclass[12pt]{article}

\usepackage{amsmath,amssymb}

\newcommand{\be}{\begin{equation}}
\newcommand{\ee}{\end{equation}}
\def\tr{{\rm tr}}
\def\cN{{\cal N}}
\def\bea{\begin{eqnarray}}
\def\eea{\end{eqnarray}}
\def\nn{\nonumber}

\def\theequation{\arabic{section}.\arabic{equation}}

\begin{document}
\begin{titlepage}
\hfill ITP-UH-09/08

\begin{center}
{\Large\bf $\cN{=}4$ superparticle and super Yang-Mills theory\\[5pt]
in $USp(4)$ harmonic superspace} \vspace{1cm}

{\large\bf I.L. Buchbinder$\,{}^{a}$, O. Lechtenfeld$\,{}^{b}$,
I.B. Samsonov$\,{}^{b,c}$\,\footnote{Alexander von Humboldt fellow at Leibniz Universit\"at Hannover.}}
\\[8pt]
\it\small a) Dept.\ of Theoretical Physics, Tomsk State
Pedagogical University, 634041 Tomsk, Russia
\\
{\tt joseph@tspu.edu.ru}\\[8pt]
b) Institut f\"ur Theoretische Physik, Leibniz Universit\"at
Hannover, 30167 Hannover, Germany\\
{\tt lechtenf@itp.uni-hannover.de}
\\[8pt]
c) Laboratory of Mathematical Physics, Tomsk Polytechnic University,
634050 Tomsk, Russia\\
{\tt samsonov@mph.phtd.tpu.edu.ru}
\end{center}
\vspace{0.5cm}

\begin{abstract}
\noindent We study the $\cN{=}4$ harmonic superparticle model,
both with and without central charge and quantize it. Since the
central charge breaks the $U(4)$ R-symmetry group of the $\cN{=}4$
superalgebra down to $USp(4)$, we consider the superparticle
dynamics in $\cN{=}4$ harmonic superspace with $USp(4)/(U(1)\times
U(1))$ harmonic variables. We show that the quantization of a
massive superparticle with central charge leads to a superfield
realization of the $\cN{=}4$ massive vector multiplet in $\cN{=}4$
harmonic superspace. In the massless case without central charge
the superparticle quantization reproduces three different
multiplets: the $\cN{=}4$ SYM multiplet, the $\cN{=}4$ gravitino
multiplet and $\cN{=}4$ supergravity multiplet.
The SYM multiplet is described by six analytic superfield
strengths with different types of analyticity. We show that these
strengths solve the $\cN{=}4$ SYM constraints and can be used for
the construction of actions in $\cN{=}4$ harmonic superspace.
\end{abstract}
\end{titlepage}

\setcounter{equation}{0}
\section{Introduction}
The $\cN{=}4$ super Yang-Mills (SYM) field theory, being the
maximally extended rigid supersymmetric model, possesses many
remarkable properties. The symmetry of this model is so large that
the only freedom in the classical action is the choice of the
gauge group, and the quantum dynamics is free of divergences. It
worth pointing out that this theory has profound relations
with superstring theory, particularly due to the AdS/CFT
correspondence (see, e.g., \cite{AdsCFT}).

The problems of $\cN{=}4$ SYM theory in the quantum domain are
mainly related to the effective action and correlation functions
of composite operators. The superfield approaches seem to be more
efficient for these purposes, since they allow one to use the
supersymmetries in explicit form. However, a description of
$\cN{=}4$ SYM theory in terms of unconstrained $\cN{=}4$
superfields is still missing. For various applications,
formulations in terms of $\cN{=}1$ superfields (see, e.g.,
\cite{N1}), in terms of $\cN{=}2$ superfields \cite{HSS,HSS1}, or
in terms of $\cN{=}3$ superfields \cite{Harm1} are used. All
attempts to find an unconstrained $\cN{=}4$ harmonic superfield
formulation for the $\cN{=}4$ SYM theory have been futile so far
\cite{N4:(,Bandos1,Howe95,Ferrara,Fer99,DHHK}, for a number of different
types of harmonic variables originating from various cosets of the
$SU(4)$ group. However, one may still hope that there exists some
other harmonic superspace, not based on some $SU(4)$ coset, which
is better suited for a superfield realization of $\cN{=}4$
supergauge theory.\footnote{As is well known, an unconstrained
superfield formulation of $\cN{=}4$ SYM theory is impossible in
standard $\cN{=}4$ superspace.} In other words, we need new
superfield representations of the known irreducible multiplets of
the $\cN{=}4$ superalgebra realized in an appropriate harmonic
superspace.

The main purpose of this paper is to construct $\cN{=}4$
superparticle models in harmonic superspace, to quantize them and
to derive superfield representations of the $\cN{=}4$ superalgebra
as a result of their quantization. It is well known that the
quantization of superparticles is closely related to superfield
formulations of the corresponding field theories. Indeed, the
superparticle models are rich of symmetries such as
reparameterization invariance, supersymmetry and, in particular
cases, the kappa-symmetry (see, e.g., \cite{superembedding} for a
review). All these symmetries are accompanied by constraints in
the Hamiltonian formulation. Upon quantization, these constraints
turn into equations of motion as well as superfield differential
constraints, which together define superfield representations of
irreducible multiplets of supersymmetry. For instance, the
quantization of the $\cN{=}1$ superparticle was achieved in
\cite{Casalbuoni,VolkovPashnev,BS,Luk1}, the $\cN{=}2$ gauge
multiplet and hypermultiplets were obtained in
\cite{Luk1,Sorokin1,Sorokin2,stv,Lus} by quantizing the $\cN{=}2$
superparticle and, finally, the $\cN{=}3$ superparticle was
recently studied by two of us \cite{BS1}, where massive and
massless $\cN{=}3$ vector multiplets as well as the $\cN{=}3$
gravitino multiplet were derived. A particular case of massless
superparticles with arbitrary $\cN{>}2$ extended supersymmetry in
$SU(\cN)$ harmonic superspace was analyzed in \cite{Sorokin1},
where the corresponding superfield strengths were derived. We
point out the significance of harmonic superparticles with
$\cN{=}2$ and $\cN{=}3$ extended supersymmetries
\cite{Sorokin1,Sorokin2,BS1},
 since they yield equations of motion for the
corresponding field theories in harmonic superspaces which possess
unconstrained superfield descriptions.

In the present paper we study models of the $\cN{=}4$ harmonic
superparticle both in the massive case with central charge and in
the massless case without central charge. It is well known that a
central charge breaks the $U(4)$ R-symmetry group of the $\cN{=}4$
superalgebra down to $USp(4)$ \cite{Fayet79,FS}. Therefore, we
consider it as crucial to introduce $USp(4)$ harmonic variables
which are employed for the corresponding $\cN{=}4$ harmonic
superspace. The various cosets of the $USp(4)$ group were
introduced and studied in \cite{IKNO}, and the corresponding
harmonic variables were further applied in
\cite{Sokatchev96,FerSok} to $d{=}5$ and $d{=}6$ $\cN{=}4$ SYM
models. In our work we find them useful also for $d{=}4$ $\cN{=}4$
SYM and superparticle models. We start with the formulation and
quantization of the $\cN{=}4$ superparticle in such a harmonic
superspace and find superfield representations of various
multiplets of the $\cN{=}4$ superalgebra. In the massive case with
nonzero central charge, the quantization leads to the massive
$\cN{=}4$ vector multiplet, represented by analytic superfields
subject to several Grassmann and harmonic shortness conditions. In
the massless case with vanishing central charge, this multiplet
reduces to the usual $\cN{=}4$ SYM multiplet if one also imposes
reality conditions. As a result, the $\cN{=}4$ SYM multiplet is
described by six analytic superfields with different types of
analyticity and harmonic shortness. Apart form the SYM multiplet,
we also derive the $\cN{=}4$ gravitino multiplet (with highest
helicity 3/2) and the multiplet of $\cN{=}4$ supergravity. These
multiplets are represented by $\cN{=}4$ chiral-analytic and chiral
superfields, respectively.

The quantization of the $\cN{=}4$ superparticle appears to be very
fruitful since it not only hints at a suitable $\cN{=}4$ harmonic
superspace based on $USp(4)$, but it also gives an appropriate
formulation of $\cN{=}4$ superfield strengths in such a harmonic
superspace. We show that exactly these superfields appear in the
solution of the $\cN{=}4$ SYM constraints with the help of
$USp(4)$ harmonic variables. Finally, we employ these superfields
for constructing some integral invariants and sketch certain
superfield actions which describe an $F^4$ term in such a harmonic
superspace. Note that similar superfields were introduced in
\cite{Bandos1,Howe95,Ferrara,Fer99,Sorokin1} by contracting the $\cN{=}4$
superfield strengths with harmonics on some coset of $SU(4)$ and
have been exploited for constructing $\cN{=}4$ invariant actions
and correlation functions of composite operators in various works
(see, e.g., \cite{DHHK,corr}).

The paper is organized as follows. In Section 2 we introduce the
$\cN{=}4$ harmonic superspace with $USp(4)$ harmonic variables and
review the basic constructions in it. In the next Section we
consider the $\cN{=}4$ harmonic superparticle model and develop
the Lagrangian and Hamiltonian formulations for it. The
quantization of the superparticle is given in Section 4, where the
superfield representations for the massive and massless vector
multiplets as well as for the gravitino and supergravity
multiplets are found. In Section 5 we show how $USp(4)$ harmonic
variables help solving the SYM constraints and construct various
actions in such an $\cN{=}4$ harmonic superspace. In a Summary we
discuss the results obtained and some ideas of their further
application to $\cN{=}4$ SYM theory. Technical details are
collected in three Appendices, where we address the problem of an
unconstrained $\cN{=}4$ superfield description of the $F^2$ term
in harmonic superspace.

\setcounter{equation}{0}
\section{$\cN{=}4$ $USp(4)$ harmonic superspace}
\subsection{$\cN{=}4$ superalgebra with central charges}
We start with a short review of $\cN{=}4$ superspace and
superalgebra constructions just to fix our notations. The
generators of $\cN{=}4$ superalgebra with central charges can be
represented by the following differential operators \be
Q^i_\alpha=\frac\partial{\partial\theta^\alpha_i}-i\bar\theta^{\dot\alpha
i} \sigma^m_{\alpha\dot\alpha}\frac\partial{\partial x^m}
 +i Z^{ij}\theta_{j\alpha},\qquad
\bar Q_{\dot\alpha
i}=-\frac\partial{\partial\bar\theta^{\dot\alpha
i}}+i\theta^\alpha_i\sigma^m_{\alpha\dot\alpha}\frac\partial{\partial
x^m} +i\bar Z_{ij}\bar\theta^j_{\dot\alpha}, \label{e1} \ee where
$\{x^m,\theta_{i\alpha},\bar\theta^i_{\dot\alpha}\}$ are the
superspace coordinates and $Z^{ij}=-Z^{ji}$, $\bar
Z_{ij}=(Z^{ij})^*$ are constant antisymmetric matrices of central
charges. The Greek letters $\alpha,\dot\alpha$ denote the $SL(2,C)$
indices while the small Latin ones $i,j,\ldots=1,2,3,4$ correspond
to R-symmetry. The operators (\ref{e1}) satisfy the standard
anticommutation relations of $\cN{=}4$ superalgebra with central
charges, \be \{Q^i_\alpha, Q^j_\beta \}
=2i\varepsilon_{\alpha\beta} Z^{ij},\quad \{\bar
Q_{i\dot\alpha},\bar Q_{j\dot\beta}
\}=-2i\varepsilon_{\dot\alpha\dot\beta} \bar Z_{ij},\quad
\{Q^i_\alpha,\bar Q_{\dot\alpha j} \}=
2i\delta^i_j\sigma^m_{\alpha\dot\alpha}\frac\partial{\partial
x^m}. \label{e4} \ee

It is well known that without central charges the $\cN{=}4$
superalgebra possesses $U(4)$ R-symmetry group. However, the
non-zero central charges break the $U(4)$ R-symmetry group down to
$USp(4)$ \cite{Fayet79,FS}. Indeed, the relations (\ref{e4}) are
invariant under those $U(4)$ transformations with the matrices
$u^i{}_k$ which leave the antisymmetric constant tensor $Z^{ij}$
invariant, $u^i{}_k u^j{}_l Z^{kl}=Z^{ij}$. Hence, $Z^{ij}$ plays
the role of invariant tensor in $Sp(4)$ group. The resulting the
R-symmetry group is given by the intersection of $U(4)$ and
$Sp(4)$ groups, that is nothing but $USp(4)$.

By applying the rotations with some unitary matrices $u^i{}_j\in
U(4)$ to the central charge tensor $Z^{ij}$ one can bring it to
the normal form \cite{FS,Zumino},
\be
Z^{ij}\to Z'{}^{ij}= u^i{}_k u^j{}_l Z^{kl} = \left(
\begin{array}{cccc}
0& -z_1&0&0\\
z_1&0&0&0\\
0&0&0&-z_2\\
0&0&z_2&0
\end{array}
 \right),
\label{e5}
\ee
where $z_1$, $z_2$ are non-negative numbers.
We restrict ourself in the further considerations to the case
$z_1=z_2\equiv z$, so we are left with the only central charge.
More generally one can consider complex central charge $z$ by
giving arbitrary phase factor to it due to the $U(1)$ rotations in
the $U(4)$ group. This is sufficient for obtaining short
representations of $\cN{=}4$ superalgebra with central charge,
when the masses of multiplets are related with the central charges
by the BPS condition \cite{Fayet79,FS},
\be m^2=z\bar z.
\label{BPS}
\ee
The central charge matrix (\ref{e5}) can be
written as
\be
Z^{ij}=z\Omega^{ij},\qquad \bar Z_{ij}=-\bar
z\Omega_{ij},
\label{e6}
\ee
where
\be
\Omega_{ij}=\left(
\begin{array}{cccc}
0& 1&0&0\\
-1&0&0&0\\
0&0&0&1\\
0&0&-1&0
\end{array}
\right),\qquad
\Omega^{ij}=(\Omega_{ij})^{-1}=\left(
\begin{array}{cccc}
0& -1&0&0\\
1&0&0&0\\
0&0&0&-1\\
0&0&1&0
\end{array}
\right).
\label{e7}
\ee
The matrix $\Omega$ will be considered further as the invariant
tensor in the $USp(4)$ group.

The $\cN{=}4$ superspace possesses the following supercovariant
Cartan forms \be \omega^M=\left\{
\begin{array}{rcl}
\omega^m&=&dx^m-id\theta_i^\alpha\sigma^m_{\alpha\dot\alpha}
\bar\theta^{i\dot\alpha}
+i\theta_i^\alpha\sigma^m_{\alpha\dot\alpha}
d\bar\theta^{i\dot\alpha},\\
\omega_i^\alpha&=&d\theta_i^\alpha,\\
\bar\omega^{i\dot\alpha}&=&d\bar\theta^{i\dot\alpha},
\end{array}
\right.
\label{e45}
\ee
which will be further used in the construction of
the superparticle Lagrangian.

\subsection{$USp(4)$ harmonic variables}
The harmonic variables on different cosets of the $USp(4)$ group
were introduced in \cite{IKNO}. In the present work we are
interested in the harmonics on the $USp(4)/(U(1)\times U(1))$
coset which we review in this subsection in some details.

The $USp(4)$ harmonic variables are $4\times4$ unitary
matrices $u=(u^i{}_j)$ preserving the antisymmetric
tensor $\Omega$ (\ref{e7}),
\be
u\in USp(4)\quad\Rightarrow\quad u\, u^\dag=1,\quad u\,\Omega\, u^{\rm T}=\Omega.
\label{e8}
\ee
Note that it is not necessary to impose the constraint
$\det u=1$ since it follows from (\ref{e8}).

Let us denote the elements of complex conjugate matrix as
$u^*=(\bar u_i{}^j)$.\footnote{We assume that the complex conjugation
flips the position of indices while the transposition changes their order,
so that $u=(u^i{}_j)$, $u^*=(\bar u_i{}^j)$, $u^{\rm T}=(u_j{}^i)$,
$u^\dag=(\bar u^j{}_i)$.}
 Then the identities (\ref{e8}) can be written
for the matrix elements as
\be
 u^i{}_j\bar u_k{}^j=\delta^i_k,
\qquad
 u^i{}_j\Omega^{jk}u^l{}_k=\Omega^{il}.
\label{e13}
\ee
As follows from (\ref{e13}),
\be
\bar u_i{}^j=\Omega_{ik}u^k{}_l\Omega^{lj},
\label{e15}
\ee
the conjugated matrix in the $USp(4)$ group is not
independent, but is expressed through the original one with the
help of invariant tensor $\Omega$. In other words, the fundamental
and conjugated representations are equivalent, similarly as for the
$SU(2)$ group. Hence, the invariant tensors $\Omega_{ij}$ and $
\Omega^{ij}$ are used to lower and rise the $USp(4)$ indices, e.g.,
\be
u^{ij}=u^i{}_k\Omega^{kj}=\Omega^{ik}\bar u_k{}^j,
\qquad
\bar u_{ij}=\Omega_{ik}u^k{}_j=\bar u_i{}^k\Omega_{kj}.
\label{e17}
\ee
Here we assume $(\Omega_{ij})^*=-\Omega^{ij}$.

Now we introduce the $usp(4)$ algebra as a space spanned on the
following differential operators
\begin{align}
&S_1=D^1_1-D^2_2,&&
  S_2=D^3_3-D^4_4,
\nn\\
&D^{(++,0)}=D^1_2,&&
  D^{(--,0)}=D^2_1,\nn\\
&D^{(0,++)}=D^3_4,&&
  D^{(0,--)}=D^4_3,\nn\\
&D^{(+,+)}=D^1_4+D^3_2,&&
  D^{(-,-)}=D^2_3+D^4_1,\nn\\
&D^{(+,-)}=D^1_3-D^4_2,&&
  D^{(-,+)}=D^2_4-D^3_1,
\label{e21}
\end{align}
where
\be
D^i_j=u^i{}_k\frac\partial{\partial u^j{}_k}.
\label{e19}
\ee
The commutation relations of the operators (\ref{e21}), given by
(\ref{e22}), show that $D^{(++,0)}$, $D^{(0,++)}$,
$D^{(+,+)}$, $D^{(-,+)}$ are rising operators,
$D^{(--,0)}$, $D^{(0,--)}$, $D^{(-,-)}$, $D^{(+,-)}$ are lowering
ones and $S_1$, $S_2$ are Cartan generators in the $USp(4)$ group.
The operators $S_1$, $S_2$ measure the $U(1)$ charges of the
generators of $USp(4)$ group,
\be
[S_1,D^{(s_1,s_2)}]=s_1D^{(s_1,s_2)},\qquad
[S_2,D^{(s_1,s_2)}]=s_2D^{(s_1,s_2)}.
\label{e23}
\ee
It is convenient to label the harmonic variables by their
$U(1)$ charges as well,
\be
u^1{}_i=u^{(+,0)}_i,\quad
u^2{}_i=u^{(-,0)}_i,\quad
u^3{}_i=u^{(0,+)}_i,\quad
u^4{}_i=u^{(0,-)}_i.
\label{e25}
\ee
The harmonic derivatives (\ref{e21}) can now be rewritten in the
more useful form for practical calculations with harmonics
(\ref{e25}),
\begin{align}
&S_1=u^{(+,0)}_i\frac\partial{\partial u^{(+,0)}_i}-
   u^{(-,0)}_i\frac\partial{\partial u^{(-,0)}_i},&&
S_2=u^{(0,+)}_i\frac\partial{\partial u^{(0,+)}_i}-
   u^{(0,-)}_i\frac\partial{\partial u^{(0,-)}_i},\nn\\
&D^{(\pm\pm,0)}=u^{(\pm,0)}_i\frac\partial{\partial u^{(\mp,0)}_i},&&
 D^{(0,\pm\pm)}=u^{(0,\pm)}_i\frac\partial{\partial
 u^{(0,\mp)}_i},\nn\\
&D^{(\pm,\pm)}=u^{(\pm,0)}_i\frac\partial{\partial u^{(0,\mp)}_i}
 +u^{(0,\pm)}_i\frac\partial{\partial u^{(\mp,0)}_i},&&
 D^{(\pm,\mp)}=u^{(\pm,0)}_i\frac\partial{\partial u^{(0,\pm)}_i}
 -u^{(0,\mp)}_i\frac\partial{\partial u^{(\mp,0)}_i}.
\label{e27}
\end{align}

Using the notations (\ref{e25}), the basic relations for
harmonics (\ref{e13}) can be written as orthogonality
\bea
&&u^{(+,0)i}u^{(-,0)}_i=u^{(0,+)i}u^{(0,-)}_i=1,
\label{e28}\\
&& u^{(+,0)}_i u^{(0,+)i}=u^{(+,0)}_i u^{(0,-)i}=
u^{(0,+)}_i u^{(-,0)i}=u^{(-,0)}_i u^{(0,-)i}=0
\label{e29}
\eea
and completeness conditions,
\be
u^{(+,0)i}u^{(-,0)}_j-u^{(+,0)}_j u^{(-,0)i}
+u^{(0,+)i}u^{(0,-)}_j-u^{(0,+)}_j u^{(0,-)i}=\delta^i_j.
\label{compl}
\ee

Apart from the usual complex conjugation there is the
following conjugation for harmonics \cite{IKNO},
\be
\widetilde{u^{(\pm,0)}_i}=u^{(0,\pm)i},\quad
\widetilde{u^{(0,\pm)}_i}=u^{(\pm,0)i},\quad
\widetilde{u^{(\pm,0)i}}=-u^{(0,\pm)}_i,\quad
\widetilde{u^{(0,\pm)i}}=-u^{(\pm,0)}_i.
\label{h-conj1}
\ee
It is the conjugation (\ref{h-conj1}) which allows one to
define real objects in harmonic superspace with $USp(4)$ harmonics.

The invariant Cartan forms on the $USp(4)$ group are given by
\bea
\omega^{(\pm\pm,0)}&=&u^{(\pm,0)i}d u^{(\pm,0)}_i,\qquad
\omega^{(0,\pm\pm)}=u^{(0,\pm)i}d u^{(0,\pm)}_i,\nn\\
\omega^{(\pm,\pm)}&=&\frac12(u^{(\pm,0)i}d u^{(0,\pm)}_i+
 u^{(0,\pm)i}d u^{(\mp,0)}_i),\nn\\
\omega^{(\pm,\mp)}&=&\frac12(u^{(0,\mp)i}d u^{(\pm,0)}_i+
 u^{(\pm,0)i}d u^{(0,\mp)}_i),\nn\\
\omega_1^{(0,0)}&=&
 \frac12(u^{(+,0)i}d u^{(-,0)}_i-u^{(-,0)}_i d u^{(+,0)i}),\nn\\
\omega_2^{(0,0)}&=&
 \frac12(u^{(0,+)i}d u^{(0,-)}_i-u^{(0,-)}_i d u^{(0,+)i}).
\label{e30} \eea These differential forms will be further used in
the construction of the superparticle Lagrangian on $\cN{=}4$
harmonic superspace with $USp(4)$ harmonic variables.

\setcounter{equation}{0}
\section{$\cN{=}4$ harmonic superparticle model}
In this section we construct the $\cN{=}4$ harmonic superparticle
model extending the conventional $\cN{=}4$ superparticle by the
Lagrangian for the harmonic variables on the coset
$USp(4)/(U(1)\times U(1))$. Then we develop the Hamiltonian
formulation and analyze the constraints. We assume everywhere that
any Lagrangian $L$ defines an action $S$ by the rule $S=\int d\tau
L$, where $\tau$ is worldline parameter. The derivatives of all
superspace variables over $\tau$ are denoted by dots, e.g., $\dot
x^m=dx^m/d\tau$, $\dot\theta_{i\alpha}=d\theta_{i\alpha}/d\tau$,
etc.

\subsection{$\cN{=}4$ superparticle Lagrangian}
The Lagrangian for a massive superparticle moving in $\cN{=}4$
superspace can be written in terms of Cartan forms (\ref{e45}) in
a standard way
\cite{Casalbuoni,VolkovPashnev,BS,Luk1},\footnote{Here we assume
that Cartan forms $\omega^M$ are pulled back
$\omega^M=\dot\omega^M(\tau) d\tau$ on the superparticle worldline
$Z^M=Z^M(\tau)$, where $Z^M$ is a set of superspace coordinates.
We hope that our notations for the Cartan forms $\omega^M$ and
their values $\dot\omega^M$ do not lead to misunderstandings. The
same concerns the Cartan forms for harmonics (\ref{e30}) and their
values appearing in the Lagrangian (\ref{e32}).} \bea
L_{\cN{=}4}&=&L_1+L_2,
\label{e46}\\
L_1&=&-\frac12(e^{-1}\dot\omega^m\dot\omega_m+em^2),
\label{e47}\\
L_2&=&-(Z^{ij}\theta_i^\alpha\dot\theta_{j\alpha}
+\bar Z_{ij}\bar\theta^i_{\dot\alpha}\dot{\bar\theta}{}^{j\dot\alpha}).
\label{e48}
\eea
Here $L_1$ is a kinetic term with the mass $m$ and $L_2$ is the
Wess-Zumino term with the central charges $Z^{ij}$, $\bar Z_{ij}$.
Further we will consider only the case when the central charges
are given by (\ref{e6}), so the Lagrangian $L_2$ takes the following form
\be
L_2=-(z\Omega^{ij}\theta_i^\alpha\dot\theta_{j\alpha}
 -\bar
 z\Omega_{ij}\bar\theta^i_{\dot\alpha}\dot{\bar\theta}{}^{j\dot\alpha}).
\label{e49}
\ee

The superparticle action is invariant under supertranslations,
\bea
\delta_\epsilon \theta_i^\alpha&=&\epsilon_i^\alpha,\qquad
\delta_\epsilon \bar\theta^{i\dot\alpha}=
 \bar\epsilon^{i\dot\alpha},\nn\\
\delta_\epsilon x^m&=&-i\epsilon_i\sigma^m\bar\theta^i
+i\theta_i\sigma^m\bar\epsilon^i,
\label{susy}
\eea
which lead to the conserved charges (supercharges),
\be
Q^i_\alpha=2ie^{-1}\dot\omega_m(\sigma^m\bar\theta^i)_\alpha
 +2z\Omega^{ij}\theta_{j\alpha},\qquad
\bar
Q_{i\dot\alpha}=-2ie^{-1}\dot\omega_m(\theta_i\sigma^m)_{\dot\alpha}
-2\bar z\Omega_{ij}\bar\theta^j_{\dot\alpha}.
\label{supercharges}
\ee
Upon quantization, the supercharges (\ref{supercharges}) turn
into the differential operators (\ref{e1}) with the superalgebra
(\ref{e4}).

If the BPS condition (\ref{BPS}) is satisfied,
the Lagrangian (\ref{e46}) respects also $\kappa$-symmetry,
\bea
\delta_\kappa\theta_{i\alpha}&=&
 -ip_m(\sigma^m\bar\kappa_i)_\alpha-\bar
 z\Omega_{ij}\kappa^j_\alpha,
\qquad
\delta_\kappa\bar\theta^i_{\dot\alpha}=
 ip_m(\kappa^i\sigma^m)_{\dot\alpha}+z\Omega^{ij}
  \bar\kappa_{j\dot\alpha},
\nn\label{e218}\\
\delta_\kappa x^m&=&i\delta_\kappa\theta_i\sigma^m\bar\theta^i
 -i\theta_i\sigma^m\delta_\kappa\bar\theta^i,
\qquad
\delta_\kappa e=-4(\bar\kappa_{i\dot\alpha}\dot{\bar\theta}{}^{i\dot\alpha}
 +\dot\theta_i^\alpha\kappa^i_\alpha),
\label{kappa}
\eea
where $\kappa^i_\alpha$, $\bar\kappa_{i\dot\alpha}$ are
anticommuting local parameters.
Despite the relation (\ref{BPS}) defines the central charge $z$ only up to a
phase, we fix this freedom without loss of generality as
\be
m=-iz=i\bar z.
\label{e96}
\ee
As it will be shown further, it is the relation (\ref{e96}) that
provides us with the correct Dirac equations for physical spinors
in massive supermultiplets.

\subsection{Momenta, constraints and Hamiltonian}
We introduce the canonical momenta for the superspace variables as
follows, \bea p_m&=&-\frac{\partial L_{\cN{=}4}}{\partial \dot
x^m}=e^{-1}\dot\omega^m,
\label{e50}\\
\pi^i_\alpha&=&\frac{\partial L_{\cN{=}4}}{\partial \dot
\theta_i^\alpha}=ip_m(\sigma^m\bar\theta^i)_\alpha+z\Omega^{ij}\theta_{j\alpha},
\label{e51}\\
\bar\pi_{i\dot\alpha}&=&\frac{\partial L_{\cN{=}4}}{
\partial \dot{\bar\theta}{}^{i\dot\alpha}}\
=ip_m(\theta_i\sigma^m)_{\dot\alpha}+\bar
z\Omega_{ij}\bar\theta^j_{\dot\alpha}=-(\pi^i_\alpha)^*.
\label{e52}
\eea
The spinorial momenta (\ref{e51},\ref{e52}) do not allow one to
express the corresponding velocities and therefore they are considered
as the constraints,
\bea
D^i_\alpha&=&-\pi^i_\alpha+ip_m(\sigma^m\bar\theta^i)_\alpha
 +z\Omega^{ij}\theta_{j\alpha}\approx 0,
\nn\\
\bar D_{i\dot\alpha}&=&\bar\pi_{i\dot\alpha}
-ip_m(\theta_i\sigma^m)_{\dot\alpha}-\bar
z\Omega_{ij}\bar\theta^j_{\dot\alpha}\approx0.
\label{e53}
\eea
As usual, the first and second class constraints are entangled in
(\ref{e53}) and their separation can be done, e.g., with the help
of space-time harmonic variables \cite{Sokathev86,Bandos2}. However, we do not
need this technique here since the correct accounting of these
constraints will be done in the next section by applying
Gupta-Bleuler quantization method in harmonic superspace.

To get the Hamiltonian for $\cN{=}4$ superparticle we perform the
Legendre transform, \be H_{\cN{=}4}=-\dot
x^mp_m+\dot\theta_i^\alpha \pi ^i_\alpha
+\dot{\bar\theta}{}^i_{\dot\alpha}\bar\pi_i^{\dot\alpha}-L_{\cN{=}4}
=-\frac e2(p^2-m^2). \label{e56} \ee The Hamiltonian is
proportional to the first-class mass-shell constraint \be
p^2-m^2\approx0 \label{e57} \ee with the Lagrange multiplier $e$.

The Poisson brackets are defined in a standard way,
\be
[x^m,p_n]_P=-\delta^m_n,\quad
\{\theta_i^\alpha,\pi^j_\beta
\}_P=-\delta_i^j\delta^\alpha_\beta,\quad
\{\bar\theta^{i\dot\alpha},\bar\pi_{j\dot\beta} \}_P
=-\delta^i_j\delta^{\dot\alpha}_{\dot\beta}.
\label{e54}
\ee
With the help of the Poisson brackets we write down the variation
for $\kappa$-symmetry transformations,
\be
\delta_\kappa=\kappa^{i\alpha}[\psi_{i\alpha},\cdot \}_P+
\bar\kappa_{i\dot\alpha}[\bar\psi^{i\dot\alpha},\cdot \}_P,
\label{e58}
\ee
where
\be
\psi_{i\alpha}=-ip_m\sigma^m{}_\alpha{}^{\dot\alpha}\bar D_{i\dot\alpha}
+\bar z\Omega_{ik} D^k_\alpha,\qquad
\bar\psi^{i\dot\alpha}=ip_m\sigma^{m\alpha\dot\alpha}D^i_\alpha
-z\Omega^{ik}\bar D_k^{\dot\alpha}
\label{e59}
\ee
are the generators of $\kappa$-transformations (\ref{kappa}).

\subsection{Lagrangian for $\cN{=}4$ harmonic superparticle}
Let us consider now the $\cN{=}4$ harmonic superspace with
$USp(4)$ harmonics,
$Z_H=\{x^m,\theta_{i\alpha},\bar\theta^i_{\dot\alpha},u \}$, where
the harmonics $u$ are $USp(4)$ matrices defined in (\ref{e8}). The
superparticle Lagrangian (\ref{e46}) should be supplemented by the
harmonic term, \be L=L_{\cN{=}4}+L_{USp(4)}, \label{e60} \ee where
\bea L_{USp(4)}&=&L_\omega+L_{WZ}+L_\lambda,
\label{e31}\\
L_\omega&=&\frac{2R^2}{e}[\dot\omega^{(++,0)}\dot\omega^{(--,0)}
+\dot\omega^{(0,++)}\dot\omega^{(0,--)}+\dot\omega^{(+,+)}\dot\omega^{(-,-)}
+\dot\omega^{(+,-)}\dot\omega^{(-,+)}],
\label{e32}\\
L_{WZ}&=&-\frac{is_1}2
  [u^{(+,0)i}\dot u^{(-,0)}_i-\dot
 u^{(+,0)i}u^{(-,0)}_i]
 -\frac{is_2}2 [u^{(0,+)i}\dot u_i^{(0,-)}
-\dot u^{(0,+)i}u^{(0,-)}_i],
\label{e33}\\
L_\lambda&=&\lambda_1(u^{(+,0)}_iu^{(-,0)i}-1)
+\lambda_2(u^{(0,+)}_i u^{(0,-)i}-1)
+\lambda^{(-,-)} u^{(+,0)}_i u^{(0,+)i}\nn\\&&
+\lambda^{(-,+)} u^{(+,0)}_i u^{(0,-)i}
+\lambda^{(+,-)} u^{(0,+)}_i u^{(-,0)i}
+\lambda^{(+,+)} u^{(-,0)}_i u^{(0,-)i}.
\label{e34}
\eea
Here $L_\omega$ is built out from the Cartan forms (\ref{e30}) and
describes the kinetic term for harmonics,
$L_{WZ}$ is the Wess-Zumino term and $L_\lambda$ takes into
account the constraints (\ref{e28},\ref{e29}) with the Lagrange
multipliers $\lambda_1$, $\lambda_2$, $\lambda^{(+,+)}$,
$\lambda^{(-,-)}$, $\lambda^{(-,+)}$, $\lambda^{(+,-)}$. $R$,
$s_1$, $s_2$ are some constants.

The main advantage of using the harmonic superspace $Z_H$ is the
possibility of passing to the harmonic projections for all objects with
$USp(4)$ indices. For instance, for the Grassmann variables we
have
\be
\theta^I_\alpha=-u^{Ii}\theta_{i\alpha},\qquad
\bar\theta^{I}_{\dot\alpha}=u^{I}{}_i\bar\theta^i_{\dot\alpha},
\label{e62}
\ee
where the index $I$ takes the following values
\be
I=\{(+,0),\ (-,0),\ (0,+),\ (0,-) \}.
\label{I}
\ee
One can promote the conjugation (\ref{h-conj1}) to such objects,
\be
\widetilde{\theta^{(\pm,0)}_\alpha}=\bar\theta^{(0,\pm)}_{\dot\alpha},\quad
\widetilde{\theta^{(0,\pm)}_\alpha}=\bar\theta^{(\pm,0)}_{\dot\alpha},
\quad
\widetilde{\bar\theta^{(0,\pm)}_{\dot\alpha}}=-\theta^{(\pm,0)}_\alpha,\quad
\widetilde{\bar\theta^{(\pm,0)}_{\dot\alpha}}=-\theta^{(0,\pm)}_\alpha.
\label{e63}
\ee

Analogously, we project the constraints (\ref{e53}) with harmonics,
\bea
D^I_\alpha&=&u^I{}_i D^i_\alpha=
 -u^I{}_i\pi^i_\alpha+ip_m(\sigma^m\bar\theta^I)_\alpha-z\theta^I_\alpha
 \approx0,\nn\\
\bar D^I_{\dot\alpha}&=&-u^{Ii}\bar D_{i\dot\alpha}=
 -u^{Ii}\bar\pi_{i\dot\alpha}-ip_m(\theta^I\sigma^m)_{\dot\alpha}
 +\bar z\bar\theta^I_{\dot\alpha}\approx0.
\label{e64}
\eea
They are also related by the conjugation as
\be
\widetilde{D^{(\pm,0)}_\alpha}=-\bar
D^{(0,\pm)}_{\dot\alpha},\quad
\widetilde{D^{(0,\pm)}_\alpha}=-\bar D^{(\pm,0)}_{\dot\alpha},
\quad
\widetilde{\bar D^{(\pm,0)}_{\dot\alpha}}=D^{(0,\pm)}_\alpha,
\quad
\widetilde{\bar D^{(0,\pm)}_{\dot\alpha}}=D^{(\pm,0)}_\alpha.
\label{e65}
\ee

Let us give here also the harmonic projections of the
$\kappa$-symmetry constraints (\ref{e59}),
\be
\psi^I_\alpha=-ip_m\sigma^m_{\alpha\dot\alpha}\bar D^{I\dot\alpha}
 +\bar z D^I_\alpha\approx0,\qquad
\bar\psi^I_{\dot\alpha}
=-ip_m\sigma^m_{\alpha\dot\alpha}D^{I\alpha}+z\bar
D^I_{\dot\alpha}\approx0,
\label{e65.1}
\ee
where index $I$ takes the values (\ref{I}). However,
apart from (\ref{e64}) and (\ref{e65.1}) there are
harmonic constraints originating from the symmetries of
the Lagrangian for harmonics (\ref{e31}). We analyze them in
details in the next subsection.

\subsection{Constraints and Hamiltonian for harmonic variables}
Let us define the canonical momenta for harmonic variables,
\bea
v^{(\pm,0)i}&=&-\frac{\partial L_{USp(4)}}{\partial\dot u^{(\mp,0)}_i}
=-\frac{R^2}e [2u^{(\mp,0)i}\dot\omega^{(\pm\pm,0)}
+u^{(0,\mp)i}\dot\omega^{(\pm,\pm)}+u^{(0,\pm)i}\dot\omega^{(\pm,\mp)}]
-\frac{is_1}2 u^{(\pm,0)i},\nn\\
v^{(0,\pm)i}&=&-\frac{\partial L_{USp(4)}}{\partial\dot u^{(0,\mp)}_i}
=-\frac{R^2}e [u^{(\mp,0)i}\dot\omega^{(\pm,\pm)}
+u^{(\pm,0)i}\dot\omega^{(\mp,\pm)}+2u^{(0,\mp)i}\dot\omega^{(0,\pm\pm)}]
-\frac{is_2}2 u^{(0,\pm)i}.\nn\\
\label{e35}
\eea
They are used in constructing of covariant harmonic momenta,
\begin{subequations}
\bea
S_1&=&u^{(+,0)}_iv^{(-,0)i}-u^{(-,0)}_i v^{(+,0)i}=-is_1,\label{e36a}\\
S_2&=&u^{(0,+)}_iv^{(0,-)i}-u^{(0,-)}_i v^{(0,+)i}=-is_2,\label{e36b}\\
D^{(\pm\pm,0)}&=&u^{(\pm,0)}_i v^{(\pm,0)i}=\mp\frac{2R^2}{e}
 \dot\omega^{(\pm\pm,0)},\label{e36c}\\
D^{(0,\pm\pm)}&=&u^{(0,\pm)}_i v^{(0,\pm)i}=\mp\frac{2R^2}{e}
 \dot\omega^{(0,\pm\pm)},\label{e36e}\\
D^{(\pm,\pm)}&=&u^{(\pm,0)}_iv^{(0,\pm)i}+u^{(0,\pm)}_i v^{(\pm,0)i}
 =\mp\frac{2R^2}{e}\dot\omega^{(\pm,\pm)},\label{e36g}\\
D^{(\pm,\mp)}&=&u^{(\pm,0)}_iv^{(0,\mp)i}-u^{(0,\mp)}_i v^{(\pm,0)i}
 =\mp\frac{2R^2}{e}\dot\omega^{(\pm,\mp)}.
\label{e36j}
\eea
\label{e36}
\end{subequations}
There are six constraints with the canonical momenta (\ref{e35})
\bea
C_1&=&u^{(+,0)}_i v^{(-,0)i}+u^{(-,0)}_i v^{(+,0)i}=0,\quad
C_2=u^{(0,+)}_i v^{(0,-)i}+u^{(0,-)}_i v^{(0,+)i}=0,\nn\\
C^{(\pm,\pm)}&=&u^{(\pm,0)}_i v^{(0,\pm)i}-u^{(0,\pm)}_i v^{(\pm,0)i}=0,\quad
C^{(\pm,\mp)}=u^{(\pm,0)}_i v^{(0,\mp)i}+u^{(0,\mp)}_i v^{(\pm,0)i}=0.
\nn\\
\label{e37}
\eea
Equations (\ref{e36},\ref{e37}) considered together allow one
to express all sixteen harmonic momenta (\ref{e35}) through
the variables $S$, $D$, $C$ in the lhs in (\ref{e36},\ref{e37}).
Therefore we will use further the covariant momenta (\ref{e36})
instead of canonical ones.

There are also the following constraints for the harmonic
variables which appear by varying (\ref{e31}) over Lagrange
multipliers,
\bea
\chi_1&=&\frac{\partial L_{USp(4)}}{\partial \lambda_1}=
 u^{(+,0)}_iu^{(-,0)i}-1=0,\quad
\chi_2=\frac{\partial L_{USp(4)}}{\partial \lambda_2}=
 u^{(0,+)}_iu^{(0,-)i}-1=0,\nn\\
\chi^{(\pm,\pm)}&=&\frac{\partial L_{USp(4)}}{\partial \lambda^{(\mp,\mp)}}
=u^{(\pm,0)}_i u^{(0,\pm)i}=0,\quad
\chi^{(\pm,\mp)}=\frac{\partial L_{USp(4)}}{\partial \lambda^{(\mp,\pm)}}
=u^{(\pm,0)}_i u^{(0,\mp)i}=0.
\label{e38}
\eea

Let us introduce the Poisson brackets for the harmonic
variables and corresponding momenta,
\be
[u^{(\pm,0)}_i,v^{(\mp,0)j}]_P=-\delta_i^j,\quad
 [u^{(0,\pm)}_i,v^{(0,\mp)j}]_P=-\delta_i^j,
\quad\mbox{other brackets vanish.}
\label{e39}
\ee
It is easy to see that the functions (\ref{e36a},\ref{e36b})
commute weakly under these brackets with the constraints
(\ref{e37},\ref{e38}) and therefore they belong to the first class
according to the Dirac's terminology. Another first-class
constraint appears as the equation of motion for the einbein
field,
\bea
0=\frac{\partial L}{\partial e}&=&
\frac12(
p^m p_m -m^2 +D^{(++,0)}D^{(--,0)}
+D^{(0,++)}D^{(0,--)}\nn\\&&+D^{(+,+)}D^{(-,-)}
+D^{(+,-)}D^{(-,+)}
 ).
\label{e39.1}
\eea
This equation is a modification of the mass-shell constraint (\ref{e57})
with the covariant momenta for the harmonic variables.

Let us now turn to the second-class constraints for the harmonic
variables. We denote the covariant momenta
(\ref{e36},\ref{e37}) and constraints (\ref{e38}) as
\bea
D^M&=&\{S_1,S_2,D^{(++,0)},D^{(--,0)},D^{(0,++)},D^{(0,--)},D^{(+,+)}
,D^{(-,-)},D^{(+,-)},D^{(-,+)} \},\nn\\
C^I&=&\{C_1,C_2,C^{(+,+)},C^{(-,-)},C^{(+,-)},C^{(-,+)} \},\nn\\
\chi_I&=&\{\chi_1,\chi_2,-\chi^{(-,-)},\chi^{(+,+)},-\chi^{(-,+)},\chi^{(+,-)} \}
\label{aaa}
\eea
and calculate their Poisson brackets,
\bea
&&[C^I,\chi_J]_P=2\delta^I_J,\qquad
[C^I,C^J]=F^{IJ}_M D^M,
\label{e40}
\eea
where $F^{IJ}_M$ is some constant matrix.
Equations (\ref{e40}) show that the constraints (\ref{e37},\ref{e38}) are
second-class and therefore should be taken into account with the help of
the Dirac bracket,
\bea
[A,B]_D&=&[A,B]_P+\frac12[A,C^I]_P[\chi_I,B]_P-\frac12[A,\chi_I]_P[C^I,B]_P
\nn\\&&
-\frac14[A,\chi_I]_PF^{IJ}_M D^M [\chi_J,B]_P,
\label{e41}
\eea
were $A$, $B$ are arbitrary two functions on a phase space for
harmonic variables.

Now we are ready to define the Hamiltonian for the particle on $USp(4)/(U(1)\times
U(1))$ coset as a Legendre transform for the Lagrangian
(\ref{e31}),
\be
H_{USp(4)}=-\dot u^{(+,0)}_i v^{(-,0)i}
-\dot u^{(-,0)}_i v^{(+,0)i}-\dot u^{(0,+)}_i v^{(0,-)i}
-\dot u^{(0,-)}_i v^{(0,+)i}-L_{USp(4)}.
\label{e42}
\ee
Expressing the velocities for harmonic variables from
(\ref{e35}) through the momenta and substituting them into
(\ref{e42}) we find
\bea
H_{USp(4)}&=&2R^2e^{-1}[\dot\omega^{(++,0)}\dot\omega^{(--,0)}
+\dot\omega^{(0,++)}\dot\omega^{(0,--)}+\dot\omega^{(+,+)}\dot\omega^{(-,-)}
+\dot\omega^{(+,-)}\dot\omega^{(-,+)}]\nn\\
&&-\omega_1^{(0,0)}(S_1-is_1)-\omega_2^{(0,0)}(S_2-is_2)-L_\lambda,
\label{e43}
\eea
where $L_\lambda$ is given by (\ref{e34}). Since the
constraints (\ref{e38}) are accounted by the Dirac bracket
(\ref{e41}), we omit $L_\lambda$ further. Note that the functions
$\omega_1^{(0,0)}$, $\omega_2^{(0,0)}$ are arbitrary, we treat
them as the Lagrange multipliers and denote further as $\mu$,
$\nu$, respectively. As a result, the Hamiltonian (\ref{e43}) is
given by
\bea
H_{USp(4)}&=&-\frac e{2R^2}[D^{(++,0)}D^{(--,0)}
+D^{(0,++)}D^{(0,--)}+D^{(+,+)}D^{(-,-)}
+D^{(+,-)}D^{(-,+)}]\nn\\
&&-\mu(S_1-is_1)-\nu(S_2-is_2).
\label{e44}
\eea

The Hamiltonian describing the dynamics of both superspace and
harmonic variables reads \be H=H_{USp(4)}+H_{\cN{=}4}, \label{e61}
\ee where $H_{\cN{=}4}$ and $H_{USp(4)}$ are given by (\ref{e56})
and (\ref{e44}), respectively.

\setcounter{equation}{0}
\section{Gupta-Bleuler quantization of massive harmonic superparticle
with central charge term}
According to the canonical quantization, one replaces the canonical
momenta (\ref{e50}--\ref{e52}), (\ref{e35}) with the following differential
operators,
\be
p_m\to i\partial_m,\quad
\pi^i_\alpha\to-i\frac\partial{\partial\theta_i^\alpha},\quad
\bar\pi_{i\dot\alpha}=-i\frac\partial{
\partial\bar\theta^{i\dot\alpha}},\quad
v^{(\pm,0)i}\to\frac\partial{\partial u^{(\mp,0)}_i},\quad
v^{(0,\pm)i}\to\frac\partial{\partial u^{(0,\mp)}_i}.
\label{e66}
\ee
The covariant harmonic momenta (\ref{e36}) turn into
the harmonic derivatives (\ref{e27}), while the Grassmann
constraints (\ref{e64}) correspond to the following covariant
spinor derivatives\footnote{The operators $D^i_\alpha$,
$\bar D_{i\dot\alpha}$ are multiplied here by $-i$ for convenience.}
\bea
D^{(\pm,0)}_\alpha&=&\pm \frac\partial{\partial\theta^{(\mp,0)\alpha}}
 +i(\sigma^m\bar\theta^{(\pm,0)})_\alpha\partial_m+iz\theta^{(\pm,0)}_\alpha,
 \nn\\
D^{(0,\pm)}_\alpha&=&\pm \frac\partial{\partial\theta^{(0,\mp)\alpha}}
 +i(\sigma^m\bar\theta^{(0,\pm)})_\alpha\partial_m+iz\theta^{(0,\pm)}_\alpha,
 \nn\\
\bar D^{(\pm,0)}_{\dot\alpha}&=&\pm \frac\partial{\partial\bar\theta^{(\mp,0)\dot\alpha}}
-i(\theta^{(\pm,0)}\sigma^m)_{\dot\alpha}\partial_m-i\bar
z\bar\theta^{(\pm,0)}_{\dot\alpha},
\nn\\
\bar D^{(0,\pm)}_{\dot\alpha}&=&\pm \frac\partial{\partial\bar\theta^{(0,\mp)\dot\alpha}}
-i(\theta^{(0,\mp)}\sigma^m)_{\dot\alpha}\partial_m-i\bar
z\bar\theta^{(0,\pm)}_{\dot\alpha}.
\label{e67}
\eea
with non-trivial anticommutation relations given by
\bea
&&\{D^{(+,0)}_\alpha,D^{(-,0)}_\beta \}=
\{D^{(0,+)}_\alpha,D^{(0,-)}_\beta
\}=-2iz\varepsilon_{\alpha\beta},\nn\\&&
\{\bar D^{(+,0)}_{\dot\alpha},\bar D^{(-,0)}_{\dot\beta} \}
=\{\bar D^{(0,+)}_{\dot\alpha},\bar D^{(0,-)}_{\dot\beta} \}
=2i\bar z\varepsilon_{\dot\alpha\dot\beta},\nn\\
&&\{D^{(+,0)}_\alpha,\bar D^{(-,0)}_{\dot\alpha} \}=
\{D^{(0,+)}_\alpha,\bar D^{(0,-)}_{\dot\alpha} \}=
-2i\sigma^m_{\alpha\dot\alpha}\partial_m.
\label{e68}
\eea

The operators (\ref{e66}) should be realized in some Hilbert space
formed by the superfunctions
\be
\Phi=\Phi(x^m,\theta_{i\alpha},\bar\theta^i_{\dot\alpha},u),
\label{e69}
\ee
which should satisfy some equations of motion
and constraints originating from the superparticle
constraints. The superparticle has both first- and second-class
constraints. The first-class
constraints (\ref{e36a},\ref{e36b},\ref{e39.1}) form closed algebra
under the Poisson or Dirac
bracket. Therefore, they all should be imposed on state
vectors,\footnote{Further we label the functions $\Phi$ by
the values of $U(1)$ charges as $\Phi^{(s_1,s_2)}$.}

\bea
&&S_1\Phi^{(s_1,s_2)}=s_1 \Phi^{(s_1,s_2)},\qquad
S_2\Phi^{(s_1,s_2)}=s_2 \Phi^{(s_1,s_2)},
\label{e70}\\
&&[\partial^m\partial_m
-\frac1{R^2}X+m^2]\Phi^{(s_1,s_2)}=0,
\label{e71}
\eea
where \footnote{Here we use a particular ordering of the harmonic
derivatives although other orderings are also possible.}
\be
X=D^{(--,0)}D^{(++,0)}+D^{(0,--)} D^{(0,++)}
+ D^{(-,-)} D^{(+,+)}+D^{(+,-)}D^{(-,+)}.
\label{e72}
\ee

The second-class constraints should be accounted either by
constructing the corresponding Dirac bracket or by applying
Gupta-Bleuler method. In our case the second-class harmonic
constraints (\ref{e37},\ref{e38}) are taken into account by the Dirac
bracket (\ref{e41}), while the spinorial ones (\ref{e64}) should be
accounted \`a la Gupta-Bleuler. It means that they have to be
divided into two complex conjugate subsets with weakly
commuting constraints in each subset.
As follows from the algebra (\ref{e68}), there are two
ways of separating the derivatives (\ref{e67}) into such subsets:
\bea
&&\{D^{(+,0)}_\alpha,D^{(0,+)}_\alpha,
\bar D^{(+,0)}_{\dot\alpha},\bar D^{(0,+)}_{\dot\alpha} \}\ \cup\
\{D^{(-,0)}_\alpha,D^{(0,-)}_\alpha,
\bar D^{(-,0)}_{\dot\alpha},\bar D^{(0,-)}_{\dot\alpha} \},
\label{e75a}\\
&&\{D^{(+,0)}_\alpha,D^{(0,-)}_\alpha,
\bar D^{(+,0)}_{\dot\alpha},\bar D^{(0,-)}_{\dot\alpha} \}\ \cup \
\{D^{(-,0)}_\alpha,D^{(0,+)}_\alpha,
\bar D^{(-,0)}_{\dot\alpha},\bar D^{(0,+)}_{\dot\alpha} \}.
\label{e75b}
\eea
Both these lines (\ref{e75a},\ref{e75b}) lead to the equivalent
superfield realizations of supersymmetry. Therefore, we consider
in details only (\ref{e75a}) and give short comments on the second case
(\ref{e75b}) in the end of this section on a particular example.
Therefore we require the superfield $\Phi^{(s_1,s_2)}$ to be
analytic,
\be
D^{(+,0)}_\alpha
\Phi^{(s_1,s_2)}=D^{(0,+)}_\alpha\Phi^{(s_1,s_2)}=
\bar D^{(+,0)}_{\dot\alpha}\Phi^{(s_1,s_2)}=\bar
D^{(0,+)}_{\dot\alpha}\Phi^{(s_1,s_2)}=0.
\label{e76}
\ee

Note that the physical states should respect
the first-class $\kappa$-symmetry constraints (\ref{e65.1})
which are not accounted so far. The problem is that the harmonic
part of the superparticle Lagrangian (\ref{e31}) violates the
$\kappa$-symmetry (\ref{kappa}). As follows from (\ref{e71}), the
states $\Phi^{(s_1,s_2)}$ acquire additional masses due to the
eigenvalues of the operator $X$ and the BPS condition (\ref{BPS})
is violated. This is not surprising as we consider a
superparticle in the harmonic superspace $Z_H=\{x^m,\theta_{i\alpha},\bar\theta^i_{\dot\alpha},u
\}$, where the harmonic variables have non-trivial dynamics
rather than playing auxiliary role.
To resolve this problem and to obtain the physical
states describing irreducible representations of supersymmetry
algebra we have to ``freeze'' the harmonic dynamics by imposing
additional harmonic constraints on the superfield
$\Phi^{(s_1,s_2)}$. Such constraints should be first-class
and should be compatible with the second-class constraints
(\ref{e76}). Since the harmonic derivatives
$D^{(++,0)}$, $D^{(0,++)}$, $D^{(+,+)}$, $D^{(-,+)}$,
$D^{(+,-)}$ leave the set of Grassmann derivatives in (\ref{e76})
invariant, we require them to annihilate the state,
\be
D^{(++,0)}\Phi^{(s_1,s_2)}=D^{(0,++)}\Phi^{(s_1,s_2)}
=D^{(+,+)}\Phi^{(s_1,s_2)}=D^{(-,+)}\Phi^{(s_1,s_2)}
=D^{(+,-)}\Phi^{(s_1,s_2)}=0.
\label{e73}
\ee
It is easy to see that the state under constraints (\ref{e73})
belongs to the kernel of the operator $X$, i.e.\
$X\,\Phi^{(s_1,s_2)}=0$. Therefore the mass-shell constraint
(\ref{e71}) reduces to
\be
(\partial^m\partial_m+m^2)\Phi^{(s_1,s_2)}=0,
\label{e74}
\ee
that is nothing but the usual Klein-Gordon equation. The BPS
condition (\ref{BPS}) is now restored and the state respects the
constraints of $\kappa$-symmetry which eliminate the unphysical degrees
of freedom.

Upon quantization, the generators of $\kappa$-symmetry
(\ref{e65.1}) turn into the differential operators,
\be
\psi^I_\alpha=\sigma^m_{\alpha\dot\alpha}\partial_m\bar D^{I\dot\alpha}
 +\bar z D^I_\alpha,\qquad
\bar\psi^I_{\dot\alpha}
=\sigma^m_{\alpha\dot\alpha}\partial_m D^{I\alpha}+z\bar
D^I_{\dot\alpha}.
\label{e77}
\ee
Owing to the analyticity (\ref{e76}), we have to impose only the
following constraints,
\bea
(\sigma^m_{\alpha\dot\alpha}\partial_m\bar D^{(-,0)\dot\alpha}
 +\bar z D^{(-,0)}_\alpha)\Phi^{(s_1,s_2)}&=&0,\nn\\
 (\sigma^m_{\alpha\dot\alpha}\partial_m\bar D^{(0,-)\dot\alpha}
 +\bar z D^{(0,-)}_\alpha)\Phi^{(s_1,s_2)}&=&0,\nn\\
(\sigma^m_{\alpha\dot\alpha}\partial_m D^{(-,0)\alpha}+z\bar
D^{(-,0)}_{\dot\alpha})\Phi^{(s_1,s_2)}&=&0,\nn\\
(\sigma^m_{\alpha\dot\alpha}\partial_m D^{(0,-)\alpha}+z\bar
D^{(0,-)}_{\dot\alpha})\Phi^{(s_1,s_2)}&=&0.
\label{e78}
\eea

The relations (\ref{e78}) can be brought to more useful form. For
this purpose we introduce the operators
\bea
Y^{(--,0)}&=&\frac i4(z\bar D^{(-,0)}_{\dot\alpha}\bar
D^{(-,0)\dot\alpha}
-\bar z D^{(-,0)\alpha}D^{(-,0)}_\alpha),\nn\\
Y^{(0,--)}&=&\frac i4(z\bar D^{(0,-)}_{\dot\alpha}\bar D^{(0,-)\dot\alpha}
-\bar z D^{(0,-)\alpha}D^{(0,-)}_\alpha),
\label{e79}
\eea
which have the following commutators with (\ref{e77})
\bea
\frac1z[D^{(+,0)}_\alpha, Y^{(--,0)}]&=&
 \sigma^m_{\alpha\dot\alpha}\partial_m \bar D^{(-,0)\dot\alpha}
 +\bar z D^{(-,0)}_\alpha,\nn\\
\frac1z[D^{(0,+)}_\alpha, Y^{(0,--)}]&=&
 \sigma^m_{\alpha\dot\alpha}\partial_m \bar D^{(0,-)\dot\alpha}
 +\bar z D^{(0,-)}_\alpha,\nn\\
-\frac1{\bar z}[\bar D^{(+,0)}_{\dot\alpha},Y^{(--,0)}]&=&
\sigma^m_{\alpha\dot\alpha}\partial_m
 D^{(-,0)\alpha}+z\bar D^{(-,0)}_{\dot\alpha},\nn\\
 -\frac1{\bar z}[\bar D^{(0,+)}_{\dot\alpha},Y^{(0,--)}]&=&
\sigma^m_{\alpha\dot\alpha}\partial_m
 D^{(0,-)\alpha}+z\bar D^{(0,-)}_{\dot\alpha}.
\label{e80}
\eea
Therefore instead of (\ref{e78}) we impose the following
first-class constraints on the superfield,
\bea
(z\bar D^{(-,0)}_{\dot\alpha}\bar
D^{(-,0)\dot\alpha}
-\bar z D^{(-,0)\alpha}D^{(-,0)}_\alpha)\Phi^{(s_1,s_2)}&=&0,\nn\\
(z\bar D^{(0,-)}_{\dot\alpha}\bar D^{(0,-)\dot\alpha} -\bar z
D^{(0,-)\alpha}D^{(0,-)}_\alpha)\Phi^{(s_1,s_2)}&=&0. \label{e81}
\eea In fact, the constraints (\ref{e81}) are stronger than
(\ref{e78}). Nevertheless, as is argued in
\cite{Sorokin1,Sorokin2} for $\cN{=}2$ superparticle, such
first-class constraints should be imposed on the state since they
appear as functions of second-class constraints (\ref{e64}).
Moreover, they can be considered as the generators of symmetries
of the superparticle Lagrangian as is shown in the Appendix 2 for
the massless case.

Let us summarize all the equations for the superfield
$\Phi^{(s_1,s_2)}$ in a single list
\be
\begin{array}l
S_1\Phi^{(s_1,s_2)}=s_1 \Phi^{(s_1,s_2)},\qquad
S_2\Phi^{(s_1,s_2)}=s_2 \Phi^{(s_1,s_2)},\\
D^{(+,0)}_\alpha
\Phi^{(s_1,s_2)}=D^{(0,+)}_\alpha\Phi^{(s_1,s_2)}=
\bar D^{(+,0)}_{\dot\alpha}\Phi^{(s_1,s_2)}=\bar
D^{(0,+)}_{\dot\alpha}\Phi^{(s_1,s_2)}=0,\\
D^{(++,0)}\Phi^{(s_1,s_2)}=D^{(0,++)}\Phi^{(s_1,s_2)}
=D^{(+,+)}\Phi^{(s_1,s_2)}=0,\\
D^{(-,+)}\Phi^{(s_1,s_2)}=D^{(+,-)}\Phi^{(s_1,s_2)}=0,\\
(z\bar D^{(-,0)}_{\dot\alpha}\bar
D^{(-,0)\dot\alpha}
-\bar z D^{(-,0)\alpha}D^{(-,0)}_\alpha)\Phi^{(s_1,s_2)}=0,\\
(z\bar D^{(0,-)}_{\dot\alpha}\bar D^{(0,-)\dot\alpha}
-\bar z D^{(0,-)\alpha}D^{(0,-)}_\alpha)\Phi^{(s_1,s_2)}=0,\\
(\partial^m\partial_m+m^2)\Phi^{(s_1,s_2)}=0.
\end{array}
\label{e82}
\ee
Note that due to the algebra (\ref{e22}) the derivatives
$D^{(+,-)}$, $D^{(-,+)}$ commute as
$[D^{(+,-)},D^{(-,+)}]=S_2-S_1$. Hence, the operator $S_2-S_1$
also annihilates the state, $(S_2-S_1)\Phi^{(s_1,s_2)}=0$ and the
resulting superfield has equal $U(1)$ charges,
$s_1=s_2$.

We point out that the constraints $D^{(-,+)}\Phi^{(s_1,s_2)}
=D^{(+,-)}\Phi^{(s_1,s_2)}=0$ restrict effectively the superfield
to depend on $USp(4)/(SU(2)\times U(1))$ harmonic variables rather
than the ones on the $USp(4)/(U(1)\times U(1))$ coset. Therefore
these constraints can be effectively resolved by considering the
$USp(4)/(SU(2)\times U(1))$ harmonics. Nevertheless, in the
present work we do not follow this way and work only with
the $USp(4)/(U(1)\times U(1))$ harmonic variables introduced
above.

\subsection{$\cN{=}4$ massive vector multiplet}
\label{sec}
Now we consider particular examples of superfields
$\Phi^{(s_1,s_2)}$ with the lowest values
of $U(1)$ charges satisfying (\ref{e82}). It is easy to see that the case $s_1=s_2=0$ is
trivial since such chargeless superfield is just a constant,
$\Phi^{(0,0)}=const$. Therefore the first physically interesting
example appears when $s_1=s_2=1$,
\be
\Phi^{(1,1)}\equiv W^{(+,+)}.
\label{e83}
\ee
One can check that for this superfield the equations in the last three
lines in (\ref{e82}) follow from the other ones, while the
relations in the first line are satisfied automatically. As a
result, the superfield (\ref{e83}) obeys
\bea
&&D^{(+,0)}_\alpha
W^{(+,+)}=D^{(0,+)}_\alpha W^{(+,+)}=
\bar D^{(+,0)}_{\dot\alpha}W^{(+,+)}=\bar
D^{(0,+)}_{\dot\alpha}W^{(+,+)}=0,
\label{e85}\\
&&D^{(++,0)}W^{(+,+)}=D^{(0,++)}W^{(+,+)}
=D^{(+,+)}W^{(+,+)}=0,
\label{e84}\\
&&D^{(-,+)}W^{(+,+)}
=D^{(+,-)}W^{(+,+)}=0.
\label{e84_}
\eea

Equations (\ref{e85}) mean that the superfield $W^{(+,+)}$ is
analytic. To resolve these constraints we first make the change of
coordinates \be
x_A^m=x^m-i\theta^{(-,0)}\sigma^m\bar\theta^{(+,0)}
-i\theta^{(+,0)}\sigma^m\bar\theta^{(-,0)}
-i\theta^{(0,-)}\sigma^m\bar\theta^{(0,+)}
-i\theta^{(0,+)}\sigma^m\bar\theta^{(0,-)}, \label{e86} \ee and
then pass from $\tau$ to $\lambda$ frame,\footnote{Here we use the
terminology of \cite{Book} used for similar constructions in
$\cN{=}2$ harmonic superspace. In fact, such passing from $\tau$
to $\lambda$ frame is nothing but the use of two different
representations for Grassmann and harmonic derivatives which are
related by the unitary operator $e^{\cal Z}$.} \bea
D^I_\alpha&\to&{\cal D}^I_\alpha=e^{\cal Z} D^I_\alpha e^{-{\cal
Z}}
 =D^I_\alpha-(D^I_\alpha {\cal Z}),\nn\\
\bar D^I_{\dot\alpha}&\to&\bar {\cal D}^I_{\dot\alpha}
=e^{\cal Z} \bar D^I_{\dot\alpha} e^{-{\cal Z}}
=\bar D^I_{\dot\alpha}-(\bar D^I_{\dot\alpha}{\cal Z}),\nn\\
W^{(+,+)}&\to&{\cal W}^{(+,+)}=e^{\cal Z} W^{(+,+)},
\label{e87}
\eea
where
\be
{\cal Z}=iz\theta^{(+,0)\alpha}\theta^{(-,0)}_\alpha
+iz\theta^{(0,+)\alpha}\theta^{(0,-)}_\alpha
+i\bar z\bar\theta^{(+,0)}_{\dot\alpha}\bar\theta^{(-,0)\dot\alpha}
+i\bar
z\bar\theta^{(0,+)}_{\dot\alpha}\bar\theta^{(0,-)\dot\alpha}
\label{e88}
\ee
is a bridge superfield.
In the analytic coordinates (\ref{e86}) the derivatives (\ref{e67}) read
\bea
{\cal D}^{(+,0)}_\alpha&=&\frac\partial{\partial\theta^{(-,0)\alpha}},
\quad
{\cal D}^{(-,0)}_\alpha=- \frac\partial{\partial\theta^{(0,+)\alpha}}
 +2i(\sigma^m\bar\theta^{(-,0)})_\alpha\partial_m+2iz\theta^{(-,0)}_\alpha,
\nn\\
{\cal D}^{(0,+)}_\alpha&=& \frac\partial{\partial\theta^{(0,-)\alpha}},
\quad
{\cal D}^{(0,-)}_\alpha=- \frac\partial{\partial\theta^{(0,+)\alpha}}
 +2i(\sigma^m\bar\theta^{(0,-)})_\alpha\partial_m+2iz\theta^{(0,-)}_\alpha,
\nn\\
\bar {\cal D}^{(+,0)}_{\dot\alpha}&=& \frac\partial{\partial\bar\theta^{(-,0)\dot\alpha}},
\quad
\bar{\cal D}^{(-,0)}_{\dot\alpha}=- \frac\partial{\partial\bar\theta^{(+,0)\dot\alpha}}
-2i(\theta^{(-,0)}\sigma^m)\partial_m-2i\bar z\bar\theta^{(-,0)}_{\dot\alpha},
\nn\\
\bar{\cal D}^{(0,+)}_{\dot\alpha}&=& \frac\partial{\partial\bar\theta^{(0,-)\dot\alpha}}
,\quad
\bar{\cal D}^{(0,-)}_{\dot\alpha}=- \frac\partial{\partial\bar\theta^{(0,+)\dot\alpha}}
-2i(\theta^{(0,-)}\sigma^m)\partial_m-2i\bar
z\bar\theta^{(0,-)}_{\dot\alpha}.
\label{e89}
\eea
Since the derivatives ${\cal D}^{(+,0)}_\alpha$, ${\cal
D}^{(0,+)}_\alpha$, $\bar {\cal D}^{(+,0)}_{\dot\alpha}$,
$\bar {\cal D}^{(0,+)}_{\dot\alpha}$ are short, the superfield ${\cal
W}^{(+,+)}$ depends only on the analytic coordinates,
\be
{\cal W}^{(+,+)}={\cal W}^{(+,+)}(x^m_A,\theta^{(+,0)}_\alpha,
\theta^{(0,+)}_\alpha,\bar\theta^{(+,0)}_{\dot\alpha},
\bar\theta^{(0,+)}_{\dot\alpha},u).
\label{e90}
\ee

To solve the constraints (\ref{e84},\ref{e84_}) we rewrite the covariant
harmonic derivatives in the analytic coordinates in $\lambda$
frame (omitting the terms vanishing on the analytic superfields),
\bea
{\cal D}^{(++,0)}&=&D^{(++,0)}-2i(\theta^{(+,0)}\sigma^m\bar\theta^{(+,0)})
\frac\partial{\partial x^m_A}
-iz(\theta^{(+,0)})^2-i\bar z(\bar\theta^{(+,0)})^2,
\nn\\
{\cal D}^{(0,++)}&=&D^{(0,++)}-2i(\theta^{(0,+)}\sigma^m\bar\theta^{(0,+)})
\frac\partial{\partial x^m_A}
-iz(\theta^{(0,+)})^2-i\bar z(\bar\theta^{(0,+)})^2,
\nn\\
{\cal D}^{(+,+)}&=&D^{(+,+)}-2i(\theta^{(+,0)}\sigma^m\bar\theta^{(0,+)}
+\theta^{(0,+)}\sigma^m\bar\theta^{(+,0)})
\frac\partial{\partial x^m_A}\nn\\&&
-2iz(\theta^{(+,0)}\theta^{(0,+)})-2i\bar z(\bar\theta^{(+,0)}\bar\theta^{(0,+)}),
\nn\\
{\cal D}^{(-,+)}&=&D^{(-,+)}-\theta^{(0,+)}_\alpha\frac\partial{
\partial\theta^{(+,0)}_\alpha}
-\bar\theta^{(0,+)}_{\dot\alpha}\frac\partial{
\partial\bar\theta^{(+,0)}_{\dot\alpha}},\nn\\
{\cal D}^{(+,-)}&=&D^{(+,-)}+\theta^{(+,0)}_\alpha\frac\partial{
\partial\theta^{(0,+)}_\alpha}
+\bar\theta^{(+,0)}_{\dot\alpha}\frac\partial{
\partial\bar\theta^{(0,+)}_{\dot\alpha}}.
\label{e91}
\eea
The full component structure of the superfield (\ref{e90}) is very
long. However, on the equations of motion (\ref{e84},\ref{e84_}) all
auxiliary fields vanish and the component decomposition
appears pretty short,
\bea
{\cal W}^{(+,+)}&=&u^{(+,0)}_{[i}u^{(0,+)}_{j]}f^{ij}
+\theta^{(+,0)\alpha}\psi^i_\alpha u^{(0,+)}_i
-\theta^{(0,+)\alpha}\psi^i_\alpha u^{(+,0)}_i
\nn\\&&
+\bar\theta^{(+,0)}_{\dot\alpha}\bar\chi^{i\dot\alpha}
 u^{(0,+)}_i
-\bar\theta^{(0,+)}_{\dot\alpha}\bar\chi^{i\dot\alpha} u^{(+,0)}_i
\nn\\&&
+iz(\theta^{(+,0)})^2f^{ij}u^{(-,0)}_{[i}u^{(0,+)}_{j]}
+iz(\theta^{(0,+)})^2f^{ij}u^{(+,0)}_{[i}u^{(0,-)}_{j]}
\nn\\&&
+i\bar z(\bar\theta^{(+,0)})^2 f^{ij}u^{(-,0)}_{[i}u^{(0,+)}_{j]}
+i\bar z(\bar\theta^{(0,+)})^2 f^{ij}u^{(+,0)}_{[i}u^{(0,-)}_{j]}
\nn\\&&
+(iz\theta^{(+,0)}\theta^{(0,+)}+i\bar z\bar\theta^{(+,0)}\bar\theta^{(0,+)})
 f^{ij}(u^{(+,0)}_{[i}u^{(-,0)}_{j]}-u^{(0,+)}_{[i}u^{(0,-)}_{j]})
\nn\\&&
+2i\theta^{(+,0)}\sigma^m\bar\theta^{(+,0)}\partial_m
 f^{ij}u^{(-,0)}_{[i}u^{(0,+)}_{j]}
+2i\theta^{(0,+)}\sigma^m\bar\theta^{(0,+)}\partial_m
 f^{ij}u^{(+,0)}_{[i}u^{(0,-)}_{j]}
\nn\\&&
+i(\theta^{(+,0)}_\alpha\bar\theta^{(0,+)}_{\dot\alpha}
+\theta^{(0,+)}_\alpha\bar\theta^{(+,0)}_{\dot\alpha})
 \sigma^{m\alpha\dot\alpha}\partial_m f^{ij}
 (u^{(+,0)}_{[i}u^{(-,0)}_{j]}-u^{(0,+)}_{[i}u^{(0,-)}_{j]})
\nn\\&&
+\theta^{(+,0)}_\alpha \theta^{(0,+)}_\beta F^{\alpha\beta}
+\bar\theta^{(+,0)}_{\dot\alpha} \bar\theta^{(0,+)}_{\dot\beta}
 \bar G^{\dot\alpha\dot\beta}
+\theta^{(+,0)}_\alpha\bar\theta^{(0,+)}_{\dot\alpha}\bar A^{\alpha\dot\alpha}
+\theta^{(0,+)}_\alpha\bar\theta^{(+,0)}_{\dot\alpha} A^{\alpha\dot\alpha}
\nn\\&&
-iz(\theta^{(+,0)})^2\theta^{(0,+)\alpha} \psi^i_\alpha
u^{(-,0)}_i
-iz(\theta^{(+,0)})^2\bar\theta^{(0,+)}_{\dot\alpha}
 \bar\chi^{i\dot\alpha} u^{(-,0)}_i
\nn\\&&
+iz(\theta^{(0,+)})^2\theta^{(+,0)\alpha}\psi^i_\alpha u^{(0,-)}_i
+iz(\theta^{(0,+)})^2\bar\theta^{(+,0)}_{\dot\alpha}
 \bar\chi^{i\dot\alpha}u^{(0,-)}_i
\nn\\&&
-i\bar z(\bar\theta^{(+,0)})^2\theta^{(0,+)\alpha}\psi^i_\alpha
 u^{(-,0)}_i
-i\bar z(\bar\theta^{(+,0)})^2\bar\theta^{(0,+)}_{\dot\alpha}
 \bar\chi^{i\dot\alpha}u^{(0,-)}_i
\nn\\&&
+i\bar z(\bar\theta^{(0,+)})^2\theta^{(+,0)\alpha}\psi^i_\alpha
u^{(0,-)}_i
+i\bar z(\bar\theta^{(0,+)})^2\bar\theta^{(+,0)}_{\dot\alpha}
 \bar\chi^{i\dot\alpha} u^{(0,-)}_i
\nn\\&&
+2i\theta^{(+,0)\alpha}\theta^{(0,+)\beta}\bar\theta^{(+,0)\dot\alpha}
 \sigma^m_{\alpha\dot\alpha}\partial_m\psi^i_\beta u^{(-,0)}_i
+2i\theta^{(+,0)\beta}\theta^{(0,+)\alpha}\bar\theta^{(0,+)\dot\alpha}
 \sigma^m_{\alpha\dot\alpha}\partial_m\psi^i_\beta u^{(0,-)}_i
\nn\\&&
+2i\bar\theta^{(+,0)\dot\alpha}\bar\theta^{(0,+)\dot\beta}
 \theta^{(+,0)\alpha}\sigma^m_{\alpha\dot\alpha}\partial_m
  \bar\chi^i_{\dot\beta}u^{(-,0)}_i
+2i\bar\theta^{(+,0)\dot\beta}\bar\theta^{(0,+)\dot\alpha}
 \theta^{(0,+)\alpha} \sigma^m_{\alpha\dot\alpha}\partial_m
  \bar\chi^i_{\dot\beta}u^{(0,-)}_i
\nn\\&&
-z^2(\theta^{(+,0)})^2(\theta^{(0,+)})^2
 f^{ij}u^{(-,0)}_{[i}u^{(0,-)}_{j]}
-\bar z^2(\bar\theta^{(+,0)})^2(\bar\theta^{(0,+)})^2
 f^{ij}u^{(-,0)}_{[i}u^{(0,-)}_{j]}
\nn\\&&
-z\bar z[(\theta^{(+,0)})^2(\bar\theta^{(0,+)})^2+
(\theta^{(0,+)})^2(\bar\theta^{(+,0)})^2] f^{ij}
 u^{(-,0)}_{[i}u^{(0,-)}_{j]}
\nn\\&&
-2z[(\theta^{(+,0)})^2\theta^{(0,+)}_\alpha\bar\theta^{(0,+)}_{\dot\alpha}
+(\theta^{(0,+)})^2\theta^{(+,0)}_\alpha\bar\theta^{(+,0)}_{\dot\alpha}]
\sigma^{m\alpha\dot\alpha}\partial_m f^{ij}  u^{(-,0)}_{[i}u^{(0,-)}_{j]}
\nn\\&&
-2\bar z
[(\bar\theta^{(+,0)})^2\theta^{(0,+)}_\alpha\bar\theta^{(0,+)}_{\dot\alpha}
+(\bar\theta^{(0,+)})^2\theta^{(+,0)}_\alpha\bar\theta^{(+,0)}_{\dot\alpha}]
\sigma^{m\alpha\dot\alpha}\partial_m f^{ij}  u^{(-,0)}_{[i}u^{(0,-)}_{j]}
\nn\\&&
+4\theta^{(+,0)}_\alpha\theta^{(0,+)}_\beta
\bar\theta^{(+,0)}_{\dot\alpha}\bar\theta^{(0,+)}_{\dot\beta}
\sigma^{m\alpha\dot\alpha}\sigma^{n\beta\dot\beta}
 \partial_m\partial_n f^{ij}  u^{(-,0)}_{[i}u^{(0,-)}_{j]}.
\label{e93}
\eea

The scalar fields in (\ref{e93}) satisfy Klein-Gordon equation,
\be
(\square+z\bar z)f^{ij}=0,
\label{e94}
\ee
while for the spinors we have Dirac equations with the correct signs
in the mass terms owing to (\ref{e96}),
\be
i\sigma^m_{\alpha\dot\alpha}\partial_m\psi^{i\alpha}-iz\bar\chi^{i\dot\alpha}=0,
\qquad
i\sigma^m_{\alpha\dot\alpha}\partial_m\bar\chi^{i\dot\beta}-i\bar
z\psi^i_\alpha=0.
\label{e95}
\ee
The fields $F_{\alpha\beta}$, $A_{\alpha\dot\alpha}$ and $\bar
G_{\dot\alpha\dot\beta}$, $\bar A_{\alpha\dot\alpha}$ are related
to each other as
\bea
&&i\sigma^{m\alpha}{}_{\dot\alpha}\partial_m F_{\alpha\beta}
+izA_{\beta\dot\alpha}=0,\qquad
i\sigma^m{}_\alpha{}^{\dot\alpha}\partial_m A_{\beta\dot\alpha}
 +i\bar z F_{\alpha\beta}=0,\nn\\
&&i\sigma^m{}_\alpha{}^{\dot\alpha}\partial_m \bar G_{\dot\alpha\dot\beta}
 +i\bar z\bar A_{\alpha\dot\beta}=0,\qquad
i\sigma^{m\alpha}{}_{\dot\alpha}\partial_m \bar A_{\alpha\dot\beta}
+i z\bar G_{\dot\alpha\dot\beta}=0,\nn\\
&&\sigma^m_{\alpha\dot\alpha}\partial_m
A^{\alpha\dot\alpha}=0,\qquad
\sigma^m_{\alpha\dot\alpha}\partial_m
\bar A^{\alpha\dot\alpha}=0.
\label{e97}
\eea
Equations (\ref{e97}) mean that the fields
$A_{\alpha\dot\alpha}$ and $\bar A_{\alpha\dot\alpha}$ obey the
relations
\bea
&&(\square+z\bar z)A_{\alpha\dot\alpha}=0,\qquad
\sigma^m_{\alpha\dot\alpha}\partial_m A^{\alpha\dot\alpha}=0,\nn\\
&& (\square+z\bar z)\bar A_{\alpha\dot\alpha}=0,\qquad
\sigma^m_{\alpha\dot\alpha}\partial_m\bar A^{\alpha\dot\alpha}=0
\label{e98} \eea and, hence, describe complex massive vector
field. We conclude that (\ref{e93}) represents a superfield
realization of $\cN{=}4$ massive vector multiplet, \bea
(\square+m^2)f^{ij}=0\quad (\Omega_{ij}f^{ij}=0)&\quad &\mbox{5
complex
scalars,}\nn\\
\begin{array}r
i\sigma^m_{\alpha\dot\alpha}\partial_m\psi^{i\alpha}+m\bar\chi^{i\dot\alpha}=0\\
i\sigma^m_{\alpha\dot\alpha}\partial_m\bar\chi^{i\dot\beta}-m\psi^i_\alpha=0
\end{array}&&
\mbox{4 Dirac spinors,}\nn\\
\begin{array}l
(\square+m^2)A_{\alpha\dot\alpha}=0,\quad
\sigma^m_{\alpha\dot\alpha}\partial_m A_{\alpha\dot\alpha}=0\\
(\square+m^2)\bar A_{\alpha\dot\alpha}=0,\quad
\sigma^m_{\alpha\dot\alpha}\partial_m\bar A_{\alpha\dot\alpha}=0
\end{array}&&\mbox{1 massive complex vector.}
\label{e98.1}
\eea

Let us now consider in short the second type of separation of
superparticle constraints (\ref{e75b}) on the examples of
a superfield $W^{(+,-)}$. The Grassmann derivatives in
(\ref{e75b}) should annihilate this superfield,
\be
D^{(+,0)}_\alpha W^{(+,-)}=D^{(0,-)}_\alpha W^{(+,-)}=
\bar D^{(+,0)}_{\dot\alpha}W^{(+,-)} = \bar D^{(0,-)}_{\dot\alpha}
W^{(+,-)}=0.
\label{e98.2}
\ee
We impose also the following harmonic constraints
\be
D^{(++,0)}W^{(+,-)}=D^{(0,--)}W^{(+,-)}=D^{(+,-)}W^{(+,-)}=
D^{(-,-)}W^{(+,-)}=D^{(+,+)}W^{(+,-)}=0,
\label{e98.3}
\ee
since harmonic derivatives in (\ref{e98.3}) leave the set of the
constraints (\ref{e98.2}) invariant.

To solve the constraints (\ref{e98.2}) we consider this superfield
in analytic coordinates,
\be
{x'}_A^m=x^m-i\theta^{(-,0)}\sigma^m\bar\theta^{(+,0)}
-i\theta^{(+,0)}\sigma^m\bar\theta^{(-,0)}
+i\theta^{(0,-)}\sigma^m\bar\theta^{(0,+)}
+i\theta^{(0,+)}\sigma^m\bar\theta^{(0,-)}
\label{e98.4}
\ee
and pass from $\tau$ to $\lambda$ frame by the rules (\ref{e87})
with the bridge
\be
{\cal Z}=iz\theta^{(+,0)\alpha}\theta^{(-,0)}_\alpha
-iz\theta^{(0,+)\alpha}\theta^{(0,-)}_\alpha
+i\bar z\bar\theta^{(+,0)}_{\dot\alpha}\bar\theta^{(-,0)\dot\alpha}
-i\bar z\bar\theta^{(0,+)}_{\dot\alpha}\bar\theta^{(0,-)\dot\alpha}.
\label{e98.5}
\ee
The component structure of the superfield ${\cal W}^{(+,-)}$ is
found in a similar way,
\bea
{\cal W}^{(+,-)}&=&u^{(+,0)}_{[i}u^{(0,-)}_{j]}f^{ij}
+\theta^{(+,0)\alpha}\psi^i_\alpha u^{(0,-)}_i
-\theta^{(0,-)\alpha}\psi^i_\alpha u^{(+,0)}_i
+\bar\theta^{(+,0)}_{\dot\alpha}\bar\chi^{i\dot\alpha}
 u^{(0,-)}_i
-\bar\theta^{(0,-)}_{\dot\alpha}\bar\chi^{i\dot\alpha} u^{(+,0)}_i
\nn\\&&
+iz(\theta^{(+,0)})^2f^{ij}u^{(-,0)}_{[i}u^{(0,-)}_{j]}
-iz(\theta^{(0,-)})^2f^{ij}u^{(+,0)}_{[i}u^{(0,+)}_{j]}
\nn\\&&
+i\bar z(\bar\theta^{(+,0)})^2 f^{ij}u^{(-,0)}_{[i}u^{(0,-)}_{j]}
-i\bar z(\bar\theta^{(0,-)})^2 f^{ij}u^{(+,0)}_{[i}u^{(0,+)}_{j]}
\nn\\&&
+(iz\theta^{(+,0)}\theta^{(0,-)}+i\bar z\bar\theta^{(+,0)}\bar\theta^{(0,-)})
 f^{ij}(u^{(+,0)}_{[i}u^{(-,0)}_{j]}-u^{(0,+)}_{[i}u^{(0,-)}_{j]})
\nn\\&&
+2i\theta^{(+,0)}\sigma^m\bar\theta^{(+,0)}\partial_m
 f^{ij}u^{(-,0)}_{[i}u^{(0,-)}_{j]}
-2i\theta^{(0,-)}\sigma^m\bar\theta^{(0,-)}\partial_m
 f^{ij}u^{(+,0)}_{[i}u^{(0,+)}_{j]}
\nn\\&&
+i(\theta^{(+,0)}_\alpha\bar\theta^{(0,-)}_{\dot\alpha}
+\theta^{(0,-)}_\alpha\bar\theta^{(+,0)}_{\dot\alpha})
 \sigma^{m\alpha\dot\alpha}\partial_m f^{ij}
 (u^{(+,0)}_{[i}u^{(-,0)}_{j]}-u^{(0,+)}_{[i}u^{(0,-)}_{j]})
\nn\\&&
+\theta^{(+,0)}_\alpha \theta^{(0,-)}_\beta F^{\alpha\beta}
+\bar\theta^{(+,0)}_{\dot\alpha} \bar\theta^{(0,-)}_{\dot\beta}
 \bar G^{\dot\alpha\dot\beta}
+\theta^{(+,0)}_\alpha\bar\theta^{(0,-)}_{\dot\alpha}\bar A^{\alpha\dot\alpha}
+\theta^{(0,-)}_\alpha\bar\theta^{(+,0)}_{\dot\alpha} A^{\alpha\dot\alpha}
\nn\\&&
+\ldots,
\label{e98.6}
\eea
where dots stand for the terms with $\theta$'s to the third and
fourth powers. All components here depend on ${x'}^m_A$ and
satisfy the free equations of motion (\ref{e98.1}).

Analogously, one can consider the superfield $W^{(-,+)}$
constrained by
\be
\begin{array}c
D^{(-,0)}_\alpha W^{(+,-)}=D^{(0,+)}_\alpha W^{(+,-)}
\bar D^{(-,0)}_{\dot\alpha}W^{(+,-)} = \bar D^{(0,+)}_{\dot\alpha}
W^{(+,-)}=0,\\
D^{(--,0)}W^{(+,-)}=D^{(0,++)}W^{(+,-)}=D^{(-,+)}W^{(+,-)}=
D^{(-,-)}W^{(+,-)}=D^{(+,+)}W^{(+,-)}=0.
\end{array}
\label{e98.8}
\ee
The component structure of $W^{(-,+)}$ is similar to
(\ref{e98.6}), but the values of the $U(1)$ charges are swapped.

\setcounter{equation}{0}
\section{Gupta-Bleuler quantization of massless superparticle}
We turn to the massless $\cN{=}4$ superparticle by considering the
model (\ref{e46}) in the massless limit, i.e. \be m=z=\bar z=0.
\label{e99} \ee The Lagrangian (\ref{e46}) reduces to \be
L_1=-\frac12e^{-1}\dot\omega^m\dot\omega_m. \label{e100} \ee We
point out that the massless superparticle has $U(4)$ R-symmetry
group rather than $USp(4)$. Therefore, it is naturally to extend
this model with the $SU(4)$ harmonics. This case was already
studied in \cite{Sorokin1}, where the superfield description of
$\cN{=}4$ vector multiplet was given. Therefore in the present
work we study the $\cN{=}4$ massless superparticle extended by
$USp(4)$ harmonic variables. Exactly this feature allows us to get
some new insight on the problem of superfield formulation of
$\cN{=}4$ SYM model in harmonic superspace.

The Lagrangian of $\cN{=}4$ harmonic superparticle reads \be
L_{\cN{=}4}=L_1+L_{USp(4)}, \label{e101} \ee where $L_1$ and
$L_{USp(4)}$ are defined in (\ref{e100}) and (\ref{e31}),
respectively. The further quantization of this model is
straightforward and can be read from the above considerations in
the limit (\ref{e99}). Here we mention only the new features.

In the massless case the algebra of covariant spinor derivatives
(\ref{e67}) has non-trivial anticommutation relations given
only in the last line in (\ref{e68}). Therefore, in the
Gupta-Bleuler quantization approach, there are eight ways of
separation of corresponding constraints,
\begin{subequations}
\bea
&&\{D^{(0,+)}_\alpha,D^{(0,-)}_\alpha,
\bar D^{(+,0)}_{\dot\alpha},\bar D^{(-,0)}_{\dot\alpha} \}\ \cup\
\{ D^{(+,0)}_{\alpha}, D^{(-,0)}_{\alpha},
\bar D^{(0,+)}_{\dot\alpha},\bar D^{(0,-)}_{\dot\alpha} \},
\label{e102a1}\\
&&\{D^{(+,0)}_\alpha,D^{(0,+)}_\alpha,
\bar D^{(+,0)}_{\dot\alpha},\bar D^{(0,+)}_{\dot\alpha} \}\ \cup\
\{D^{(-,0)}_\alpha,D^{(0,-)}_\alpha,
\bar D^{(-,0)}_{\dot\alpha},\bar D^{(0,-)}_{\dot\alpha} \},
\label{e102b}\\
&&\{D^{(+,0)}_\alpha,D^{(0,-)}_\alpha,
\bar D^{(+,0)}_{\dot\alpha},\bar D^{(0,-)}_{\dot\alpha} \}\ \cup\
\{D^{(-,0)}_\alpha,D^{(0,+)}_\alpha,
\bar D^{(-,0)}_{\dot\alpha},\bar D^{(0,+)}_{\dot\alpha} \},
\label{e102c}\\
&&\{D^{(+,0)}_\alpha,\bar D^{(0,-)}_{\dot\alpha},
 D^{(-,0)}_{\alpha}, D^{(0,-)}_{\alpha} \}\ \cup\
\{\bar D^{(+,0)}_{\dot\alpha},\bar D^{(0,+)}_{\dot\alpha},
\bar D^{(-,0)}_{\dot\alpha}, D^{(0,+)}_{\alpha} \},
\label{e102f}\\
&&\{D^{(+,0)}_\alpha,D^{(0,+)}_\alpha,
 D^{(-,0)}_{\alpha},\bar D^{(0,+)}_{\dot\alpha} \}\ \cup\
\{\bar D^{(+,0)}_{\dot\alpha},D^{(0,-)}_\alpha,
\bar D^{(-,0)}_{\dot\alpha},\bar D^{(0,-)}_{\dot\alpha} \},
\label{e102d}\\
&&\{D^{(+,0)}_\alpha,D^{(0,+)}_\alpha,
\bar D^{(+,0)}_{\dot\alpha}, D^{(0,-)}_{\alpha} \}\ \cup\
\{ D^{(-,0)}_{\alpha},\bar D^{(0,+)}_{\dot\alpha},
\bar D^{(-,0)}_{\dot\alpha},\bar D^{(0,-)}_{\dot\alpha} \},
\label{e102e}\\
&&\{\bar D^{(-,0)}_{\dot\alpha},D^{(0,+)}_\alpha,
 D^{(-,0)}_{\alpha}, D^{(0,-)}_{\alpha} \}\ \cup\
\{\bar D^{(+,0)}_{\dot\alpha},\bar D^{(0,+)}_{\dot\alpha},
 D^{(+,0)}_{\alpha},\bar D^{(0,-)}_{\dot\alpha} \},
\label{e102g}\\
&&\{D^{(+,0)}_\alpha,D^{(0,+)}_\alpha,
D^{(-,0)}_{\alpha}, D^{(0,-)}_{\alpha} \}\ \cup\
\{\bar D^{(+,0)}_{\dot\alpha},\bar D^{(0,+)}_{\dot\alpha},
\bar D^{(-,0)}_{\dot\alpha},\bar D^{(0,-)}_{\dot\alpha} \}.
\label{e102a}
\eea
\label{e102}
\end{subequations}
As before, we assume that the state is described by a superfield
$\Phi^{(s_1,s_2)}$ subject to (\ref{e70}). To take into
account the constraints (\ref{e102}) we claim
\be
\{D_{\alpha,\dot\alpha}\} \Phi^{(s_1,s_2)}=0,
\label{e102.1}
\ee
where $\{D_{\alpha,\dot\alpha}\}$ are covariant spinor derivatives from one of the
subsets in (\ref{e102}). We have to impose also
the harmonic constraints,
\be
\{D^A\} \Phi^{(s_1,s_2)}=0,
\label{e102.2}
\ee
where $\{D^A\}$ is a subset of harmonic derivatives (\ref{e27}) which
leave the set of Grassmann derivatives $\{D_{\alpha,\dot\alpha}
\}$ invariant. Apart from these constraints there are also
$\kappa$-symmetry ones with the generators (\ref{e77}). In the
massless case they lead to the following equations
\be
\sigma^m_{\alpha\dot\alpha}\partial_m\bar D^{I\dot\alpha}
\Phi^{(s_1,s_2)}=0,\qquad
\sigma^m_{\alpha\dot\alpha}\partial_m D^{I\alpha}
\Phi^{(s_1,s_2)}=0.
\label{e102.3}
\ee
Using the algebra of covariant spinor derivatives
one can show that the constraints (\ref{e102.3}) follow from
\be
\bar D^I_{\dot\alpha}\bar D^{J\dot\alpha}
\Phi^{(s_1,s_2)}=0,\qquad
D^{I\alpha} D^J_\alpha
\Phi^{(s_1,s_2)}=0,
\label{e102.4}
\ee
where the indices $I,J$ have the values (\ref{I}). As explained in
the Appendix 2, the constraints (\ref{e102.4}) originate from the
bosonic version of $\kappa$-symmetry. We point out that not all
the constraints (\ref{e102.2},\ref{e102.4}) are independent,
however they are all first-class. Some of them become trivial when
definite subset in (\ref{e102}) is chosen.

As a result, all the first-class and one half of second-class
constraints are taken into account by the equations
(\ref{e102.1})--(\ref{e102.4}) for the superfield
$\Phi^{(s_1,s_2)}$. In the following subsections we consider the
particular examples of massless representations of $\cN{=}4$
superalgebra on such superfields with lowest values of $U(1)$
charges $s_1$, $s_2$, which correspond to different ways of
separations of constraints (\ref{e102}) and describe different
multiplets.

\subsection{Supergauge multiplet}
Regarding to the separations of constraints
(\ref{e102a1})--(\ref{e102c}), there are two essentially different
superfield realizations of $\cN{=}4$ supergauge multiplets. In the
next subsections we consider both of them separately.

\subsubsection{Chargeless superfield representation}
\label{chargeless}
According to the separation of constraints (\ref{e102a1}), the
superfield $\Phi^{(s_1,s_2)}$ should be annihilated by the
following Grassmann derivatives
\be
D^{(0,+)}_\alpha \Phi^{(s_1,s_2)}=
D^{(0,-)}_\alpha \Phi^{(s_1,s_2)}=
\bar D^{(+,0)}_{\dot\alpha} \Phi^{(s_1,s_2)}=
\bar D^{(-,0)}_{\dot\alpha} \Phi^{(s_1,s_2)}=0.
\label{x1}
\ee
As follows from (\ref{e102.4}), the linearity conditions read
\bea
&&(D^{(+,0)})^2 \Phi^{(s_1,s_2)}=(D^{(-,0)})^2 \Phi^{(s_1,s_2)}
=(D^{(+,0)}D^{(-,0)}) \Phi^{(s_1,s_2)}=0,\nn\\
&&(\bar D^{(0,+)})^2  \Phi^{(s_1,s_2)}=
(\bar D^{(0,-)})^2  \Phi^{(s_1,s_2)}=
(\bar D^{(0,+)}\bar D^{(0,-)})\Phi^{(s_1,s_2)}=0.
\label{x1.1}
\eea

We have to impose also the harmonic constraints, to fix the
dynamics over the harmonic variables. Only the operators
$D^{(++,0)}$, $D^{(--,0)}$, $D^{(0,++)}$, $D^{(0,--)}$ leave the
set of Grassmann derivatives in (\ref{x1}) invariant. Therefore we
claim
\be
D^{(++,0)}\Phi^{(s_1,s_2)}=D^{(--,0)}\Phi^{(s_1,s_2)}=D^{(0,++)}\Phi^{(s_1,s_2)}
=D^{(0,--)}\Phi^{(s_1,s_2)}=0.
\label{x2}
\ee
According to the algebra (\ref{e22}),
$[D^{(++,0)},D^{(--,0)}]=S_1$, $[D^{(0,++)},D^{(0,++)}]=S_2$,
the constraints (\ref{x2}) imply $s_1=s_2=0$. Hence, the
state under considerations is realized by chargeless superfield
$\Phi^{(0,0)}\equiv W_1$.

However, the constraints (\ref{x2}) do not fix the
harmonic dynamics completely. Therefore we impose also the
quadratic harmonic constraint,
\be
D^{(+,+)}D^{(+,+)}W_1=0.
\label{x3}
\ee
As a consequence of the algebra (\ref{e22}), the operators
$D^{(-,-)}D^{(+,+)}$, $D^{(+,-)}D^{(-,+)}$ annihilate the state
as well and the whole operator (\ref{e72}) vanishes on this superfield,
\be
X\,W_1=0.
\label{x4}
\ee
Therefore we are convinced that the constraints
(\ref{x2},\ref{x3}) are sufficient to fix the dynamics over the
harmonic variables and to eliminate all unphysical degrees of
freedom in this superfield.

Let us rewrite all these constraints for the superfield $W_1$ in a single
list,
\be
\begin{array}l
D^{(0,+)}_\alpha W_1=
D^{(0,-)}_\alpha W_1=
\bar D^{(+,0)}_{\dot\alpha} W_1=
\bar D^{(-,0)}_{\dot\alpha} W_1=0,\\
D^{(++,0)}W_1=D^{(--,0)}W_1=D^{(0,++)}W_1
=D^{(0,--)}W_1=0,\\
D^{(+,+)}D^{(+,+)}W_1=0.
\end{array}
\label{x5}
\ee
The linearity constraints (\ref{x1.1}) are not in this list since
they are not independent but follow from (\ref{x5}).

To solve the constraints in the first line of (\ref{x5}) we pass
to the analytic coordinates,
\be
x_A^m=x^m-i\theta^{(0,-)}\sigma^m\bar\theta^{(0,+)}
+i\theta^{(0,+)}\sigma^m\bar\theta^{(0,-)}
-i\theta^{(+,0)}\sigma^m\bar\theta^{(-,0)}
+i\theta^{(-,0)}\sigma^m\bar\theta^{(+,0)},
\label{x6}
\ee
in which the Grassmann derivatives in (\ref{x1}) become
short,
\be
D^{(0,\pm)}_\alpha=\pm\frac\partial{\partial\theta^{(0,\mp)\alpha}},\qquad
\bar D^{(\pm,0)}_{\dot\alpha}=\pm\frac\partial{\partial\bar\theta^{
(\mp,0)\dot\alpha}}.
\label{x7}
\ee
Therefore the superfield $W_1$ depends only on one half of
Grassmann variables,
\be
W_1=W_1(x_A^m,\theta^{(+,0)}_\alpha,\theta^{(-,0)}_\alpha,
\bar\theta^{(0,+)}_{\dot\alpha},\bar\theta^{(0,-)}_{\dot\alpha},u).
\label{x8}
\ee
Let us rewrite also the harmonic derivatives in (\ref{x5}) in such
analytic coordinates (we omit the terms acting trivially on $W_1$),
\bea
D_A^{(\pm\pm,0)}&=&D^{(\pm\pm,0)}+\theta^{(\pm,0)}\frac\partial{
\partial\theta^{(\mp,0)}},\qquad
D_A^{(0,\pm\pm)}=D^{(0,\pm\pm)}+\theta^{(0,\pm)}\frac\partial{
\partial\theta^{(0,\mp)}},
\label{x12}\\
D_A^{(+,+)}&=&D^{(+,+)}-2i(\theta^{(+,0)}\sigma^m\bar\theta^{(0,+)}
-\theta^{(0,+)}\sigma^m\bar\theta^{(+,0)})\partial_m\nn\\&&
+\theta^{(0,+)}\frac\partial{\partial\theta^{(-,0)}}
+\bar\theta^{(+,0)}\frac\partial{\partial\bar\theta^{(0,-)}}.
\label{x13}
\eea
Since the operators (\ref{x12}) do not contain the
spatial derivatives, the constraints in the second line of
(\ref{x5}) are not kinematical but purely algebraical for the
components of $W_1$. In fact, these constraints show that $W_1$
depends effectively on $USp(4)/(SU(2)\times SU(2))$ harmonic
variables since the derivatives in (\ref{x2}) form two $su(2)$
algebras. In principle, such harmonic constraints (\ref{x2}) can
be solved manifestly by choosing the appropriate harmonic
variables. As a result, only the constraint in the last line of
(\ref{x5}) is true equation of motion for the superfield
$W_1$.

Finally, we impose also the reality constraint,
\be
\widetilde{W_1}=W_1,
\label{x14}
\ee
where the conjugation is defined in (\ref{h-conj1}).

Now we give the solution of all the constraints for the superfield
$W_1$,
\bea
W_1&=&\phi+if^{ij}(u^{(+,0)}_{[i}u^{(-,0)}_{j]}-u^{(0,+)}_{[i}u^{(0,-)}_{j]})
\nn\\&&
+i\theta^{(+,0)\alpha}\psi^i_\alpha u^{(-,0)}_i
-i\theta^{(-,0)\alpha}\psi^i_\alpha u^{(+,0)}_i
+i\bar\theta^{(0,+)}_{\dot\alpha}\bar\psi^{i\dot\alpha} u^{(0,-)}_i
-i\bar\theta^{(0,-)}_{\dot\alpha}\bar\psi^{i\dot\alpha} u^{(0,+)}_i
\nn\\&&
+\theta^{(+,0)}_\alpha\theta^{(-,0)}_\beta F^{(\alpha\beta)}
+\bar\theta^{(0,+)}_{\dot\alpha}\bar\theta^{(0,-)}_{\dot\beta}
 \bar F^{(\dot\alpha\dot\beta)}
\nn\\&&
+4\theta^{(+,0)}_\alpha\bar\theta^{(0,+)}_{\dot\alpha}
\partial_{\alpha\dot\alpha}f^{ij}u^{(-,0)}_{[i}u^{(0,-)}_{j]}
+4\theta^{(-,0)}_\alpha\bar\theta^{(0,-)}_{\dot\alpha}
\partial_{\alpha\dot\alpha}f^{ij}u^{(+,0)}_{[i}u^{(0,+)}_{j]}
\nn\\&&
-4\theta^{(+,0)}_\alpha\bar\theta^{(0,-)}_{\dot\alpha}
\partial_{\alpha\dot\alpha}f^{ij}u^{(-,0)}_{[i}u^{(0,+)}_{j]}
-4\theta^{(-,0)}_\alpha\bar\theta^{(0,+)}_{\dot\alpha}
\partial_{\alpha\dot\alpha}f^{ij}u^{(+,0)}_{[i}u^{(0,-)}_{j]}
\nn\\&&
-2\theta^{(+,0)\alpha}\theta^{(-,0)\beta}\bar\theta^{(0,+)\dot\alpha}
\partial_{(\alpha\dot\alpha}\psi^i_{\beta)} u^{(0,-)}_i
+2\theta^{(+,0)\alpha}\theta^{(-,0)\beta}\bar\theta^{(0,-)\dot\alpha}
\partial_{(\alpha\dot\alpha}\psi^i_{\beta)} u^{(0,+)}_i
\nn\\&&
-2\theta^{(+,0)\alpha}\bar\theta^{(0,+)\dot\beta}\bar\theta^{(0,-)\dot\alpha}
\partial_{\alpha(\dot\alpha}\bar\psi^i_{\dot\beta)} u^{(-,0)}_i
+2\theta^{(-,0)\alpha}\bar\theta^{(0,+)\dot\beta}\bar\theta^{(0,-)\dot\alpha}
\partial_{\alpha(\dot\alpha}\bar\psi^i_{\dot\beta)} u^{(+,0)}_i
\nn\\&&
+4\theta^{(+,0)}_\alpha\theta^{(-,0)}_\beta
\bar\theta^{(0,+)}_{\dot\alpha}\bar\theta^{(0,-)}_{\dot\beta}
\partial^{\alpha\dot\alpha}\partial^{\beta\dot\beta}
[\phi-if^{ij}(u^{(+,0)}_{[i}u^{(-,0)}_{j]}-u^{(0,+)}_{[i}u^{(0,-)}_{j]})].
\label{x15}
\eea
All component fields here depend only on $x_A^m$ and satisfy the
corresponding free equations of motion,
\bea
\square \phi =0&\quad&\mbox{1 real scalar},\nn\\
\square f^{ij}=0,\quad (f^{ij}\Omega_{ij}=0)&&
\mbox{5 real scalars},\nn\\
\sigma^{m\alpha}{}_{\dot\alpha}\partial_m\psi^i_\alpha=0,\quad
\sigma^m{}_\alpha{}^{\dot\alpha}\partial_m\bar\psi^i_{\dot\alpha}=0&&
\mbox{4 Weyl spinors},\nn\\
\sigma^{m\alpha}{}_{\dot\alpha}\partial_m
F_{(\alpha\beta)}=0,\quad
\sigma^{m}{}_\alpha{}^{\dot\alpha}\partial_m \bar
F_{(\dot\alpha\dot\beta)}=0&& \mbox{1 Maxwell field} \label{x16}
\eea and reality conditions, \be \bar\phi=\phi,\quad \bar
f^{ij}=f^{ij},\quad
\overline{\psi^i_\alpha}=\bar\psi_{i\dot\alpha},\quad
\overline{F_{(\alpha\beta)}}=\bar F_{(\dot\alpha\dot\beta)}.
\label{x16.1} \ee As a result, the $\cN{=}4$ SYM multiplet is
embedded into the superfield $W_1$.

Let us consider now the Grassmann derivatives in the second subset
of (\ref{e102a1}). We denote the superfield annihilated by these
variables as $W_2$. Similarly as $W_1$, it satisfies the
following constraints,
\be
\begin{array}l
\bar D^{(0,+)}_{\dot\alpha} W_2=
\bar D^{(0,-)}_{\dot\alpha} W_2=
 D^{(+,0)}_\alpha W_2=
 D^{(-,0)}_\alpha W_2=0,\\
D^{(++,0)}W_2=D^{(--,0)}W_2=D^{(0,++)}W_2
=D^{(0,--)}W_2=0,\\
D^{(-,-)}D^{(-,-)}W_2=0,\\
\widetilde{W_2}=W_2.
\end{array}
\label{x17} \ee The Grassmann constraints in (\ref{x17}) are
solved by passing to the corresponding analytic coordinates, \be
W_2=W_2({x'}_A^m,\bar\theta^{(+,0)}_{\dot\alpha},\bar\theta^{(-,0)}_{\dot\alpha},
\theta^{(0,+)}_{\alpha},\theta^{(0,-)}_{\alpha},u), \label{x18}
\ee where \be
{x'}_A^m=x^m+i\theta^{(0,-)}\sigma^m\bar\theta^{(0,+)}
-i\theta^{(0,+)}\sigma^m\bar\theta^{(0,-)}
+i\theta^{(+,0)}\sigma^m\bar\theta^{(-,0)}
-i\theta^{(-,0)}\sigma^m\bar\theta^{(+,0)}. \label{x19} \ee The
component structure of $W_2$ is analogous to (\ref{x15}), but the
values of $U(1)$ charges of superspace coordinates should be
changed appropriately. Therefore, $W_2$ also describes the
$\cN{=}4$ SYM multiplet.

\subsubsection{Charged superfield representation}
\label{charged}
Consider the separation of constraints (\ref{e102b}) on the
example of a superfield $W^{(+,+)}$ with $U(1)$ charges
$s_1=s_2=1$. According to (\ref{e102b}) and (\ref{e73}), it
satisfies similar equations as the massive vector multiplet
considered in sect.\ \ref{sec},
\bea
&&D^{(+,0)}_\alpha
W^{(+,+)}=D^{(0,+)}_\alpha W^{(+,+)}=
\bar D^{(+,0)}_{\dot\alpha}W^{(+,+)}=\bar
D^{(0,+)}_{\dot\alpha}W^{(+,+)}=0,
\label{e115}\\
&&D^{(++,0)}W^{(+,+)}=D^{(0,++)}W^{(+,+)}
=D^{(+,+)}W^{(+,+)}=0,
\label{e114}\\
&&D^{(-,+)}W^{(+,+)}
=D^{(+,-)}W^{(+,+)}=0.
\label{e114.1}
\eea
We point out that the harmonic constraints
(\ref{e114.1}) are not kinematical, but
they just show that $W^{(+,+)}$ depends effectively on
$USp(4)/(SU(2)\times U(1))$ harmonic variables. These constraints
can be solved manifestly by the appropriate choice of harmonics,
but we prefer to work here with the harmonic variables
on the $USp(4)/(U(1)\times U(1))$ coset, introduced above.

There are also linearity conditions (\ref{e102.4}),
\be
\begin{array}c
(D^{(-,0)})^2W^{(+,+)}=(D^{(0,-)})^2W^{(+,+)}= (\bar
D^{(-,0)})^2W^{(+,+)}=
(\bar D^{(0,-)})^2W^{(+,+)}=0,\\
 D^{(-,0)}D^{(0,-)}W^{(+,+)}=
\bar D^{(-,0)}\bar D^{(0,-)}W^{(+,+)}=0.
\end{array}
\label{e115.1}
\ee
Finally, in the massless case we impose also the reality
condition,
\be
\widetilde{W^{(+,+)}}=W^{(+,+)}.
\label{e116}
\ee

The solution of these constraints can be obtained from (\ref{e93}) by
putting the central charges to zero, $z=\bar z=0$ and taking into
account the reality (\ref{e116}),
\bea
W^{(+,+)}&=&u^{(+,0)}_{[i}u^{(0,+)}_{j]}if^{ij}
+i\theta^{(+,0)\alpha}\psi^i_\alpha u^{(0,+)}_i
-i\theta^{(0,+)\alpha}\psi^i_\alpha u^{(+,0)}_i
\nn\\&&
+i\bar\theta^{(+,0)}_{\dot\alpha}\bar\psi^{i\dot\alpha}
 u^{(0,+)}_i
-i\bar\theta^{(0,+)}_{\dot\alpha}\bar\psi^{i\dot\alpha} u^{(+,0)}_i
\nn\\&&
-2\theta^{(+,0)}_\alpha\bar\theta^{(+,0)}_{\dot\alpha}
 \sigma^{m\alpha\dot\alpha}\partial_m
 f^{ij}u^{(-,0)}_{[i}u^{(0,+)}_{j]}
-2\theta^{(0,+)}_\alpha\bar\theta^{(0,+)}_{\dot\alpha}
 \sigma^{m\alpha\dot\alpha}\partial_m
 f^{ij}u^{(+,0)}_{[i}u^{(0,-)}_{j]}
\nn\\&&
-(\theta^{(+,0)}_\alpha\bar\theta^{(0,+)}_{\dot\alpha}
+\theta^{(0,+)}_\alpha\bar\theta^{(+,0)}_{\dot\alpha})
 \sigma^{m\alpha\dot\alpha}\partial_m f^{ij}
 (u^{(+,0)}_{[i}u^{(-,0)}_{j]}-u^{(0,+)}_{[i}u^{(0,-)}_{j]})
\nn\\&&
+\theta^{(+,0)}_\alpha \theta^{(0,+)}_\beta F^{(\alpha\beta)}
+\bar\theta^{(+,0)}_{\dot\alpha} \bar\theta^{(0,+)}_{\dot\beta}
 \bar F^{(\dot\alpha\dot\beta)}
\nn\\&&
+i(\theta^{(+,0)}\sigma^m\bar\theta^{(0,+)}
+\theta^{(0,+)}\sigma^m\bar\theta^{(+,0)}) \varepsilon_{mnrs}G^{nrs}
\nn\\&&
-2\theta^{(+,0)\alpha}\theta^{(0,+)\beta}\bar\theta^{(+,0)\dot\alpha}
 \sigma^m_{\alpha\dot\alpha}\partial_m\psi^i_\beta u^{(-,0)}_i
-2\theta^{(+,0)\beta}\theta^{(0,+)\alpha}\bar\theta^{(0,+)\dot\alpha}
 \sigma^m_{\alpha\dot\alpha}\partial_m\psi^i_\beta u^{(0,-)}_i
\nn\\&&
-2\bar\theta^{(+,0)\dot\alpha}\bar\theta^{(0,+)\dot\beta}
 \theta^{(+,0)\alpha}\sigma^m_{\alpha\dot\alpha}\partial_m
  \bar\psi^i_{\dot\beta}u^{(-,0)}_i
-2\bar\theta^{(+,0)\dot\beta}\bar\theta^{(0,+)\dot\alpha}
 \theta^{(0,+)\alpha} \sigma^m_{\alpha\dot\alpha}\partial_m
  \bar\psi^i_{\dot\beta}u^{(0,-)}_i
\nn\\&&
+4i\theta^{(+,0)}_\alpha\theta^{(0,+)}_\beta
\bar\theta^{(+,0)}_{\dot\alpha}\bar\theta^{(0,+)}_{\dot\beta}
\sigma^{m\alpha\dot\alpha}\sigma^{n\beta\dot\beta}
 \partial_m\partial_n f^{ij}  u^{(-,0)}_{[i}u^{(0,-)}_{j]},
\label{e117}
\eea
where all components depend on the analytic coordinates
(\ref{e86}). Owing to the reality condition (\ref{e116}) we have
\be
\bar f^{ij}=f^{ij},\quad
\overline{\psi^i_{\alpha}}=\bar\psi_{i\dot\alpha},\quad
\overline{F_{(\alpha\beta)}}=\bar F_{(\dot\alpha\dot\beta)},\quad
\overline{G^{mnr}}=G^{mnr}.
\label{e118}
\ee
These components satisfy free equations of motion,
\bea
\square f^{ij}=0\quad (f^{ij}\Omega_{ij}=0)&\quad&\mbox{5 real scalars,}
\label{e119}\\
\sigma^{m\alpha}{}_{\dot\alpha}\partial_m \psi^i_{\alpha}=0,\quad
\sigma^m{}_\alpha{}^{\dot\alpha}\partial_m\bar\psi^i_{\dot\alpha}=0
&&\mbox{4 Weyl spinors,}
\label{e120}\\
\sigma^{m\alpha}{}_{\dot\alpha}\partial_m F_{(\alpha\beta)}=0,\quad
\sigma^m{}_\alpha{}^{\dot\alpha}\partial_m \bar
F_{(\dot\alpha\dot\beta)}=0&&\mbox{1 real Maxwell field,}
\label{e121}\\
\partial_m G^{mnr}=0,\quad \varepsilon_{mnrs}\partial^m G^{nrs}=0&&
\mbox{1 antisymmetric tensor field.}
\label{e122}
\eea
It is well known that the antisymmetric tensor field (\ref{e122})
is classically equivalent to the scalar field on-shell. Indeed, introducing the
field
\be
L_m=\frac16\varepsilon_{mnrs}G^{nrs},
\label{e122.1}
\ee
the equations (\ref{e122}) can be rewritten as
\be
\partial^m L_m =0,\qquad \partial_m L_n-\partial_n L_m=0.
\label{e122.2}
\ee
A general solution of (\ref{e122.2}) is expressed through the scalar
field subject to the Klein-Gordon equation,
\be
L_m=\partial_m \phi,\qquad \square\phi=0.
\label{e123}
\ee

As a result, we see that the superfield (\ref{e117}) describes the
$\cN{=}4$ vector multiplet, in which one of the scalars is
dualized and represented by the antisymmetric tensor field. Here
$G^{mnr}$ is a strength of the antisymmetric tensor field which
can always be expressed through its potential as
$G^{mnr}=\partial^m B^{nr}+\partial^n B^{rm}+\partial^r B^{mn}$.

In conclusion of this subsection we comment briefly on the other
ways of separation of constraints (\ref{e102c}) on the examples of
the superfields $W^{(+,-)}$, $W^{(-,+)}$. Similarly as in the
massive case, these superfields are constrained by
(\ref{e98.2},\ref{e98.3}) and (\ref{e98.8}) respectively. The
component structure can be read from (\ref{e98.6}) by putting the
central charges to zero, $z=\bar z=0$. Therefore, these
superfields also describe the $\cN{=}4$ vector multiplet
(\ref{e119})--(\ref{e122}) in which one of the scalars is
represented by the antisymmetric tensor field.

\subsection{Gravitino multiplet}
Let us consider the separation of constraints (\ref{e102f}) and
require the superfield $\Phi^{(0,+)}$ to be annihilated by the
following Grassmann derivatives,
\be
\bar D^{(+,0)}_{\dot\alpha}\Phi^{(0,+)}=\bar
D^{(0,+)}_{\dot\alpha}\Phi^{(0,+)}=
\bar D^{(-,0)}_{\dot\alpha}\Phi^{(0,+)}= D^{(0,+)}_{\alpha}\Phi^{(0,+)}=0.
\label{e124}
\ee
The harmonic derivatives commuting with the spinor derivatives in
(\ref{e124}) are $D^{(++,0)}$, $D^{(--,0)}$, $D^{(0,++)}$, $D^{(+,+)}$,
$D^{(-,+)}$. Therefore we require them to annihilate the state,
\be
D^{(++,0)}\Phi^{(0,+)}=D^{(--,0)}\Phi^{(0,+)}=D^{(0,++)}\Phi^{(0,+)}=
D^{(+,+)}\Phi^{(0,+)}=D^{(-,+)}\Phi^{(0,+)}=0.
\label{e125}
\ee
Finally, the superfield satisfies the linearity conditions (\ref{e102.4}),
\be
\begin{array}c
(\bar D^{(0,-)})^2\Phi^{(0,+)}=(D^{(+,0)})^2\Phi^{(0,+)}=(D^{(-,0)})^2\Phi^{(0,+)}
= (D^{(0,-)})^2\Phi^{(0,+)}=0,\\
 D^{(-,0)}D^{(0,-)}\Phi^{(0,+)}=
D^{(+,0)}D^{(0,-)}\Phi^{(0,+)}= D^{(+,0)}D^{(-,0)} \Phi^{(0,+)}=0.
\end{array}
\label{e126}
\ee

To solve the constraints (\ref{e124}) we pass to the
chiral-analytic coordinates,
\be
y^m=x^m+i\theta^{(-,0)}\sigma^m \bar\theta^{(+,0)}
-i\theta^{(+,0)}\sigma^m\bar\theta^{(-,0)}
-i\theta^{(0,-)}\sigma^m\bar\theta^{(0,+)}
-i\theta^{(0,+)}\sigma^m\bar\theta^{(0,-)},
\label{e127}
\ee
in which the derivatives in (\ref{e124},\ref{e125}) are
given by (as usual, we omit here the terms acting on $\Phi^{(0,+)}$
trivially)
\bea
\bar
D^{(\pm,0)}_{\dot\alpha}&=&\pm\frac\partial{\partial\bar\theta^{(\mp,0)\dot\alpha}},
\quad
\bar D^{(0,+)}_{\dot\alpha}=\frac\partial{\partial\bar\theta^{(0,-)\dot\alpha}},
\quad
D^{(0,+)}=\frac\partial{\partial\theta^{(0,-)\alpha}},
\label{e129}\\
D_A^{(\pm\pm,0)}&=&D^{(\pm\pm,0)}
 +\theta^{(\pm,0)}_\alpha\frac\partial{\partial\theta^{(\mp,0)}_\alpha},
\quad
D_A^{(0,++)}=D^{(0,++)}-2i\theta^{(0,+)}\sigma^m\bar\theta^{(0,+)}\partial_m,
\label{e131}\\
D_A^{(\pm,+)}&=&D^{(\pm,+)}-2i\theta^{(\pm,0)}\sigma^m\bar\theta^{(0,+)}\partial_m
 \pm\theta^{(0,+)}_\alpha\frac\partial{\partial\theta^{(\mp,0)}_\alpha}.
\label{e134}
\eea
Since the derivatives (\ref{e129}) are short,
the constraints (\ref{e124}) are manifestly solved by
\be
\Phi^{(0,+)}=\Phi^{(0,+)}(y^m,\theta^{(+,0)}_\alpha,\theta^{(-,0)}_\alpha,
\theta^{(0,+)}_\alpha,\bar\theta^{(0,+)}_{\dot\alpha},u).
\label{e135}
\ee
Using the expressions for the harmonic derivatives
(\ref{e131},\ref{e134}) one can easily check that the following
component expression for $\Phi^{(0,+)}$ solves (\ref{e125}),
\bea
\Phi^{(0,+)}&=&f^i u^{(0,+)}_i
-\theta^{(+,0)\alpha}\psi^{ij}_\alpha u^{(-,0)}_{[i}u^{(0,+)}_{j]}
+\theta^{(-,0)\alpha}\psi^{ij}_\alpha u^{(+,0)}_{[i}u^{(0,+)}_{j]}
\nn\\&&
+\theta^{(0,+)\alpha}\rho_\alpha
-\frac12\theta^{(0,+)\alpha}\psi^{ij}_\alpha(u^{(+,0)}_{[i}u^{(-,0)}_{j]}-u^{(0,+)}_{[i}u^{(0,-)}_{j]})
+\bar\theta^{(0,+)}_{\dot\alpha}\bar\lambda^{\dot\alpha}
\nn\\&&
+\theta^{(+,0)}_{(\alpha}\theta^{(-,0)}_{\beta)}F^{i\alpha\beta}u^{(0,+)}_i
-\theta^{(+,0)}_{(\alpha}\theta^{(0,+)}_{\beta)}F^{i\alpha\beta}u^{(-,0)}_i
+\theta^{(-,0)}_{(\alpha}\theta^{(0,+)}_{\beta)}F^{i\alpha\beta}u^{(+,0)}_i
\nn\\&&
+2i\theta^{(+,0)}\sigma^m\bar\theta^{(0,+)}\partial_m f^i
 u^{(-,0)}_i
-2i\theta^{(-,0)}\sigma^m\bar\theta^{(0,+)}\partial_m f^i
 u^{(+,0)}_i\nn\\&&
+2i\theta^{(0,+)}\sigma^m\bar\theta^{(0,+)}\partial_m f^i
 u^{(0,-)}_i\nn\\&&
+\theta^{(+,0)\alpha}\theta^{(-,0)\beta}\theta^{(0,+)\gamma}T_{(\alpha\beta\gamma)}
-2i\theta^{(+,0)\alpha}\theta^{(-,0)\beta}\bar\theta^{(0,+)\dot\alpha}
\sigma^m_{(\alpha\dot\alpha}\partial_m\rho_{\beta)}\nn\\&&
-i\theta^{(+,0)\alpha}\theta^{(-,0)\beta}\bar\theta^{(0,+)\dot\alpha}
 \sigma^m_{(\alpha\dot\alpha}\partial_m\psi^{ij}_{\beta)}
 (u^{(+,0)}_{[i}u^{(-,0)}_{j]}-u^{(0,+)}_{[i}u^{(0,-)}_{j]})\nn\\&&
+2i\theta^{(-,0)\alpha}\theta^{(0,+)\beta}\bar\theta^{(0,+)\dot\alpha}
\sigma^m_{(\alpha\dot\alpha}\partial_m\psi^{ij}_{\beta)}u^{(+,0)}_{[i}u^{(0,-)}_{j]}
\nn\\&&
-2i\theta^{(+,0)\alpha}\theta^{(0,+)\beta}\bar\theta^{(0,+)\dot\alpha}
\sigma^m_{(\alpha\dot\alpha}\partial_m\psi^{ij}_{\beta)}u^{(-,0)}_{[i}u^{(0,-)}_{j]}
\nn\\&&
+2i\theta^{(+,0)\alpha}\theta^{(-,0)\beta}\theta^{(0,+)\gamma}\bar\theta^{(0,+)\dot\alpha}
 \sigma^m_{\gamma\dot\alpha}\partial_m
 F^i_{\alpha\beta}u^{(0,-)}_i.
\label{e136}
\eea
All the component fields in (\ref{e136}) depend on $y^m$ given by
(\ref{e127}) and satisfy free equations of motion,
\bea
\square f^{i}=0&\quad&\mbox{4 complex scalars,}
\label{e137}\nn\\
\sigma^{m\alpha}{}_{\dot\alpha}\partial_m \rho_{\alpha}=0,\quad
\sigma^m{}_\alpha{}^{\dot\alpha}\partial_m\bar\lambda_{\dot\alpha}=0
&&\mbox{2 Weyl spinors,}
\label{e138}\nn\\
\sigma^{m\alpha}{}_{\dot\alpha}\partial_m \psi^{ij}_{\alpha}=0\quad
(\psi^{ij}_\alpha\Omega_{ij}=0)
&&\mbox{5 Weyl spinors,}
\label{e139}\nn\\
\sigma^{m\alpha}{}_{\dot\alpha}\partial_m F^i_{(\alpha\beta)}=0
 &&\mbox{4 real Maxwell fields,}
\label{e140}\nn\\
\sigma^{m\alpha}{}_{\dot\alpha}\partial_m
T_{(\alpha\beta\gamma)}=0 &&\mbox{1 gravitino.} \label{e141} \eea
As a result we obtain the $\cN{=}4$ gravitino multiplet.

The other ways of separations of constraints
(\ref{e102d},\ref{e102e},\ref{e102g}) also lead to the gravitino
multiplets realized by the superfields $\Phi^{(0,-)}$,
$\Phi^{(-,0)}$, $\Phi^{(+,0)}$, respectively. The constraints for
these superfields can be easily read from the general expressions
(\ref{e102.1})--(\ref{e102.4}). The component structure is
analogous to (\ref{e136}) with appropriate change of $U(1)$
charges.

\subsection{$\cN{=}4$ supergravity multiplet}
Consider the separation of constraints (\ref{e102a}) on the
example of chargeless superfield $\Phi^{(0,0)}\equiv\Phi$. We impose the derivatives
in the second set in (\ref{e102a}) as the constraints on $\Phi$,
\be
\bar D^{(+,0)}_{\dot\alpha}\Phi=\bar D^{(0,+)}_{\dot\alpha}\Phi=
\bar D^{(-,0)}_{\dot\alpha}\Phi=\bar D^{(0,-)}_{\dot\alpha}\Phi=0,
\label{e103}
\ee
which show this superfield to be chiral,
\be
\bar D_{i\dot\alpha}\Phi=0.
\label{e103.1}
\ee
This
constraint is explicitly solved by passing to the chiral
coordinates $y^m=x^m+i\theta_i\sigma^m\bar\theta^i$,
\be
\Phi=\Phi(y^m,\theta_{i\alpha},u).
\label{e104}
\ee

Next, we impose the harmonic constraints (\ref{e73}) \be
D^{(++,0)}\Phi=D^{(0,++)}\Phi =D^{(+,+)}\Phi=D^{(-,+)}\Phi
=D^{(+,-)}\Phi=0, \label{e105} \ee which mean that $\Phi$ is
harmonic independent, \be
\Phi(y^m,\theta_{i\alpha},u)=\Phi(y^m,\theta_{i\alpha}).
\label{e106} \ee Finally, there are the linearity constraints
(\ref{e102.4}) originating from $\kappa$-symmetry (see Appendix 2
for details), \be D^{i\alpha}D^j_\alpha \Phi=0. \label{e107} \ee
It is well known that the solution of the equation (\ref{e107})
describes the $\cN{=}4$ supergravity multiplet
\cite{Fer99,N2constr}. Its component structure is given by \be
\Phi=\phi+\theta_i^\alpha\psi^i_\alpha+
\theta_i^\alpha\theta_j^\beta F^{[ij]}_{(\alpha\beta)}
+\theta_i^\alpha\theta_j^\beta\theta_k^\gamma\varepsilon^{ijkl}
T_{l(\alpha\beta\gamma)}
+\theta_i^\alpha\theta_j^\beta\theta_k^\gamma\theta_l^\delta\varepsilon^{ijkl}
C_{(\alpha\beta\gamma\delta)}, \label{e108} \ee where the
component fields satisfy \bea \square\phi=0&\qquad& \mbox{1
complex scalar,}\nn
\label{e109}\\
\sigma^{m\alpha}{}_{\dot\alpha}\partial_m \psi^i_\alpha=0
&&\mbox{4 Weyl spinors,}
\nn\label{e110}\\
\sigma^{m\alpha}{}_{\dot\alpha}\partial_m F^{[ij]}_{(\alpha\beta)}=0&&
\mbox{6 Maxwell fields,}
\nn\label{e111}\\
\sigma^{m\alpha}{}_{\dot\alpha}\partial_m
T_{i(\alpha\beta\gamma)}=0&& \mbox{4 gravitini,}
\nn\label{e112}\\
\sigma^{m\alpha}{}_{\dot\alpha}\partial_m
C_{(\alpha\beta\gamma\delta)}=0&& \mbox{1 Weyl tensor.}
\label{e113} \eea As a result, this superfield describes the
multiplet of $\cN{=}4$ supergravity.

\setcounter{equation}{0}
\section{Applications of $USp(4)/(U(1)\times U(1))$
harmonic superspace to the $\cN{=}4$ SYM model} The quantization
of $\cN{=}4$ harmonic superparticle has demonstrated an important
role of the specific $\cN{=}4$ harmonic superspace with
$USp(4)/(U(1)\times U(1))$ harmonic variables\footnote{Such
harmonic variables were introduced in \cite{IKNO}.} and the
corresponding superfields. We have seen that this superspace is
naturally associated with $\cN{=}4$ superparticle and therefore it
is instructive to study its properties and to try to develop field
theory in it.

The purpose of this section is to apply the $USp(4)/(U(1)\times
U(1))$ harmonic superspace to the $\cN{=}4$ SYM model and show
that the superfields, obtained in quantizing the superparticle,
appear naturally in the solution of $\cN{=}4$ SYM constraints. We
consider also some possibilities of constructing the invariant
actions depending on these superfields in harmonic superspace and
discuss their relevance to the $\cN{=}4$ SYM model.

\subsection{Harmonic superspace analysis of $\cN{=}4$ SYM constraints}
Let us consider standard $\cN{=}4$ superspace
$Z^M=\{x^m,\theta_{i\alpha},\bar\theta^i_{\dot\alpha}\}$ with
supercovariant spinor derivatives \be
D^i_\alpha=\frac\partial{\partial\theta^\alpha_i}+i\bar\theta^{\dot\alpha
i} \sigma^m_{\alpha\dot\alpha}\frac\partial{\partial x^m} ,\qquad
\bar D_{\dot\alpha
i}=-\frac\partial{\partial\bar\theta^{\dot\alpha
i}}-i\theta^\alpha_i\sigma^m_{\alpha\dot\alpha}\frac\partial{\partial
x^m}. \label{l1} \ee According to the generic procedure of
superspace formulation of the extended supersymmetric models
\cite{N2constr}, one introduces the gauge connections for these
derivatives, \be
D^i_\alpha\to\nabla^i_\alpha=D^i_\alpha+V^i_\alpha, \qquad \bar
D_{i\dot\alpha}\to \bar\nabla_{i\dot\alpha}= \bar
D_{i\dot\alpha}+\bar V_{i\dot\alpha} \label{l2} \ee and defines
the superfield strengths by the following anticommutators, \be
\{\nabla^i_\alpha,\nabla^j_\beta
\}=2\varepsilon_{\alpha\beta}W^{ij},\qquad
\{\bar\nabla_{i\dot\alpha},\bar\nabla_{j\dot\beta} \}=
2\varepsilon_{\dot\alpha\dot\beta}\bar W_{ij}. \label{l3} \ee It
is well known that the following $\cN{=}4$ SYM constraints put the
superfield strengths on-shell \cite{N2constr}, \bea &&\bar
D_{i\dot\alpha}W^{jk}=\frac13(\delta_i^j\bar D_{l\dot\alpha}W^{lk}
-\delta_i^k\bar D_{l\dot\alpha}W^{lj}),
\label{l5}\\
&& D^i_\alpha W^{jk}+D^j_\alpha W^{ik}=0,
\label{l6}\\
&&\overline{W^{ij}}=\bar W_{ij}=\frac12\varepsilon_{ijkl}W^{kl}.
\label{l4}
\eea

Let us project the strengths $W^{ij}$ with harmonics,
\be
W^{ij}\to W^{IJ}=u^I{}_iu^J{}_j W^{ij},
\label{l7}
\ee
where the indices $I,J$ take the values (\ref{I}). We denote these
superfields also as
\bea
W_1&=&u^{(0,+)}_i u^{(0,-)}_j W^{ij},\qquad
W_2=u^{(+,0)}_i u^{(-,0)}_j W^{ij},\nn\\
W^{(+,+)}&=&u^{(+,0)}_i u^{(0,+)}_j W^{ij},\qquad
W^{(-,-)}=u^{(-,0)}_i u^{(0,-)}_j W^{ij},\nn\\
W^{(+,-)}&=&u^{(+,0)}_i u^{(0,-)}_j W_{ij},\qquad
W^{(-,+)}=u^{(-,0)}_i u^{(0,+)}_j W_{ij}.
\label{l8}
\eea
Contracting equations (\ref{l5},\ref{l6}) with harmonics we find a
number of Grassmann analyticity constraints for these superfields,
\bea
&&D^{(0,+)}_\alpha W_1=D^{(0,-)}_\alpha W_1
=\bar D^{(+,0)}_{\dot\alpha}W_1
=\bar D^{(-,0)}_{\dot\alpha}W_1=0,\nn
\label{l9}\\
&&D^{(+,0)}_\alpha W_2=D^{(-,0)}_\alpha W_2
=\bar D^{(0,+)}_{\dot\alpha}W_2
=\bar D^{(0,-)}_{\dot\alpha}W_2=0,\nn
\label{l10}\\
&&D^{(+,0)}_\alpha W^{(+,+)}=D^{(0,+)}_\alpha W^{(+,+)}
=\bar D^{(+,0)}_{\dot\alpha}W^{(+,+)}
=\bar D^{(0,+)}_{\dot\alpha}W^{(+,+)}=0,\nn
\label{l11}\\
&&D^{(-,0)}_\alpha W^{(-,-)}=D^{(0,-)}_\alpha W^{(-,-)}
=\bar D^{(-,0)}_{\dot\alpha}W^{(-,-)}
=\bar D^{(0,-)}_{\dot\alpha}W^{(-,-)}=0,\nn
\label{l12}\\
&&D^{(+,0)}_\alpha W^{(+,-)}=D^{(0,-)}_\alpha W^{(+,-)}
=\bar D^{(+,0)}_{\dot\alpha}W^{(+,-)}
=\bar D^{(0,-)}_{\dot\alpha}W^{(+,-)}=0,\nn
\label{l13}\\
&&D^{(-,0)}_\alpha W^{(-,+)}=D^{(0,+)}_\alpha W^{(-,+)}
=\bar D^{(-,0)}_{\dot\alpha}W^{(-,+)}
=\bar D^{(0,+)}_{\dot\alpha}W^{(-,+)}=0.
\label{l14}
\eea
Moreover, by construction, the superfields (\ref{l8}) are
annihilated by the following harmonic derivatives,
\bea
&&D^{(++,0)}W_{1}=D^{(--,0)}W_{1}=D^{(0,++)}W_{1}=D^{(0,--)}W_{1}
=(D^{(+,+)})^2W_{1}=0,\nn
\label{l15}\\
&&D^{(++,0)}W_{2}=D^{(--,0)}W_{2}=D^{(0,++)}W_{2}=D^{(0,--)}W_{2}
=(D^{(-,-)})^2W_{2}=0,\nn\\
&&D^{(++,0)}W^{(+,+)}=D^{(0,++)}W^{(+,+)}=D^{(+,+)}W^{(+,+)}
=D^{(+,-)}W^{(+,+)}=D^{(-,+)}W^{(+,+)}=0,\nn
\label{l16}\\
&&D^{(--,0)}W^{(-,-)}=D^{(0,--)}W^{(-,-)}=D^{(-,-)}W^{(-,-)}
=D^{(+,-)}W^{(-,-)}=D^{(-,+)}W^{(-,-)}=0,\nn
\label{l17}\\
&&D^{(++,0)}W^{(+,-)}=D^{(0,--)}W^{(+,-)}=D^{(+,-)}W^{(+,-)}
=D^{(+,+)}W^{(+,-)}=D^{(-,-)}W^{(+,-)}=0,\nn
\label{l18}\\
&&D^{(--,0)}W^{(-,+)}=D^{(0,++)}W^{(-,+)}=D^{(-,+)}W^{(-,+)}
=D^{(+,+)}W^{(-,+)}=D^{(-,-)}W^{(-,+)}=0.\nn\\
\label{l19}
\eea
We see that the superfield strengths $W_1$, $W_2$,
$W^{(+,+)}$, $W^{(-,-)}$, $W^{(+,-)}$, $W^{(-,+)}$ introduced in the
subsections \ref{chargeless}, \ref{charged}
satisfy the same constraints (\ref{l14},\ref{l19}) and therefore
 give the solutions of $\cN{=}4$ SYM constraints
(\ref{l5},\ref{l6}).

Let us now consider the reality constraint (\ref{l4}). Applying
the following identities with harmonics
\bea
u^{(+,0)i}u^{(-,0)j}\varepsilon_{ijkl}&=&2u^{(0,+)}_{[k}u^{(0,-)}_{l]},
 \nn\\
u^{(0,+)i}u^{(0,-)j}\varepsilon_{ijkl}&=&2u^{(+,0)}_{[k}u^{(-,0)}_{l]},
 \nn\\
u^{(+,0)i}u^{(0,-)j}\varepsilon_{ijkl}&=&-2u^{(+,0)}_{[k}u^{(0,-)}_{l]},
 \nn\\
u^{(0,+)i}u^{(-,0)j}\varepsilon_{ijkl}&=&-2u^{(0,+)}_{[k}u^{(-,0)}_{l]},
 \nn\\
u^{(+,0)i}u^{(0,+)j}\varepsilon_{ijkl}&=&-2u^{(+,0)}_{[k}u^{(0,+)}_{l]},
 \nn\\
u^{(-,0)i}u^{(0,-)j}\varepsilon_{ijkl}&=&-2u^{(-,0)}_{[k}u^{(0,-)}_{l]},
\label{e119.1}
\eea
we find that (\ref{l4}) leads to the
reality properties of superfield strengths,
\be
\begin{array}c
\widetilde{W_{1,2}}=W_{1,2},\quad
\widetilde{W^{(+,+)}}=W^{(+,+)},\quad
\widetilde{W^{(-,-)}}=W^{(-,-)},\\
\widetilde{W^{(+,-)}}=W^{(-,+)},\quad
\widetilde{W^{(-,+)}}=W^{(+,-)}.
\end{array}
\label{e119.2} \ee These conjugation rules for superfield
strengths coincide with (\ref{x14},\ref{e116}) which were
previously introduced in the superparticle considerations. This
establishes the correspondence between the $\cN{=}4$ superfield
strengths (\ref{l3}) and the superfields obtained by the
superparticle quantization.

\subsection{Actions in $USp(4)/(U(1)\times U(1))$ harmonic
superspace} \label{actions} We have seen that the superfields
(\ref{l8}) satisfying the constraints and equations of motion
(\ref{l14},\ref{l19}) appear naturally in the solution of
$\cN{=}4$ SYM constraints (\ref{l5})--(\ref{l4}). Now we address a
question, whether it is possible to construct with the help of
these superfields any $\cN{=}4$ invariant superfield functionals
in $USp(4)/(U(1)\times U(1))$ harmonic superspace, which can be
treated as the actions of some $\cN{=}4$ supersymmetric field
models. For instance, similar superfields on $\cN{=}4$ harmonic
superspace with $SU(4)$ harmonic variables were used in the
construction of different integral invariants in \cite{DHHK}.

Each of the superfields (\ref{l8}) lies in its own
analytic subspace parameterized by eight Grassmann variables, as is seen in
each line in (\ref{l14}). We restrict ourself to two
superfields, $W^{(+,+)}$ and $W_1$, the others can be studied in a
similar way.

Let us consider first $W^{(+,+)}$, which lies in the analytic
subspace with the coordinates
$\{x^m_A,\theta^{(+,0)}_\alpha,\theta^{(0,+)}_\alpha, \bar
\theta^{(+,0)}_{\dot\alpha},\bar \theta^{(0,+)}_{\dot\alpha},u
\}$, where $x^m_A$ is given by (\ref{e86}). In general, one can
consider the following functional depending on this superfield \be
\int d\zeta^{(-4,-4)}{\cal F}(W^{(+,+)}), \label{e150.1} \ee where
$d\zeta^{(-4,-4)}$ is the dimesionless analytic measure defined in
(\ref{e143}). Since such a function $\cal F$ should have definite
$U(1)$ charges, $s_1=s_2=4$, it can be only quartic, \be S_4=g\int
d\zeta^{(-4,-4)}(W^{(+,+)})^4, \label{e150.2} \ee where $g$ is a
coupling constant of mass dimension $-4$. Using the component
decomposition (\ref{e117}) for $W^{(+,+)}$ it is easy to see that
the action (\ref{e150.2}) contains the following term in
components, \be S_4\sim g\int d^4x\,
F^{\alpha\beta}F_{\alpha\beta} \bar F^{\dot\alpha\dot\beta}\bar
F_{\dot\alpha\dot\beta}+\ldots, \label{e150.3} \ee where
$F_{\alpha\beta}$, $\bar F_{\dot\alpha\dot\beta}$ are the
spinorial components of the Maxwell strength and the dots indicate
the terms for all other component fields. Since the term in the
rhs of (\ref{e150.3}) appears in the fourth order of the
decomposition of the Born-Infeld action (e.g. see
\cite{Tseytlin,IZ}), one can treat the action (\ref{e150.2}) as a
part of an $\cN{=}4$ supersymmetric generalization of Born-Infeld
theory.

Consider now the chargeless superfield $W_1$ in the analytic
superspace with the coordinates
$\{x_A^m,\theta^{(+,0)}_\alpha,\theta^{(-,0)}_\alpha,
\bar\theta^{(0,+)}_{\dot\alpha},\bar\theta^{(0,-)}_{\dot\alpha},u\}$,
where $x^m_A$ is given by (\ref{x6}). In general, one can write
the following functional with it, \be \int d\zeta\,{\cal F}(W_1),
\label{e151} \ee where $d\zeta$ is the analytic measure given by
(\ref{e152a}). Since the measure in (\ref{e151}) is chargeless,
there are no restrictions on the function $\cal F$. Of particular
interest is the quartic potential, \be \int d\zeta\,(W_1)^4\sim
\int d^4x \, F^{\alpha\beta}F_{\alpha\beta} \bar
F^{\dot\alpha\dot\beta}\bar F_{\dot\alpha\dot\beta}+\ldots,
\label{e152} \ee as it may have some relations to the $\cN{=}4$
supersymmetric generalization of Born-Infeld theory as mentioned
above. One can easily check the presence of $F^4$ term in the
component structure of (\ref{e152}) by applying the component
decomposition for $W_1$ given by (\ref{x15}). Another interesting
example is given by the logarithmic potential, \be \int
d\zeta\,\ln(W_1/\Lambda)\sim \int d^4x \frac{F^4}{\phi^4},
\label{e153} \ee where $\Lambda$ is some scale which makes the
combination $W_1/\Lambda$ dimensionless. Clearly, the expression
in lhs of (\ref{e153}) is scale independent despite the manifest
presence of dimensionful parameter. In components, such functional
reproduces the first leading term in the low-energy effective
action of $\cN{=}4$ SYM model (see, e.g., \cite{Echaya}). Note
that the $\cN{=}4$ superfield description given by (\ref{e153})
for such terms is even simple than the one with $\cN{=}2$
superfields \cite{Echaya}. In the work \cite{BISZ} we noticed that
it is much more difficult to write down similar terms within
$\cN{=}3$ harmonic superspace approach.

In this subsection we saw that the superfield strengths (\ref{l8})
allowed us to construct some actions with the gauge fields in the
fourth power while the $\cN{=}4$ superfield description for $F^2$
terms remains unclear. Moreover, the strengths (\ref{l8}) are
on-shell objects constrained by (\ref{l19}). Therefore the
classical action for $\cN{=}4$ SYM model, if it admits an
$\cN{=}4$ superfield description, should be constructed in terms
of some potentials for these superfield strengths. One such
attempt is undertaken in the Appendix 3, where we introduce the
analytic prepotentials and study some superfield action with it.
However, this question requires further independent studies and
lies beyond the frame of the present work.

\section{Summary}
In this paper we have studied the construction of massive and
massless $\cN{=}4$ superparticle models in $\cN{=}4$ harmonic
superspace and their quantization. The crucial point of our
considerations is the use of $USp(4)$ harmonic variables since
exactly this group corresponds to the R-symmetry of $\cN{=}4$
superalgebra with central charge. Since the mass of the
superparticle should be equal to its central charge, as required
for the construction of BPS supermultiplets, both massive and
massless superparticles can be studied and quantized in such an
$\cN{=}4$ $USp(4)/(U(1)\times U(1))$ harmonic superspace. The
quantization leads straightforwardly to the superfield realization
of physically interesting multiplets of $\cN{=}4$ supersymmetry.
Namely, in the massive case, the $\cN{=}4$ massive vector
multiplet is described by four analytic superfields $W^{(+,+)}$,
$W^{(-,-)}$, $W^{(+,-)}$, $W^{(-,+)}$ with different types of
analyticity obeying also harmonic shortness constraints, which
serve as the equations of motion for these superfields. In the
massless case these superfields reduce to the usual $\cN{=}4$ SYM
multiplet if additional reality constraints are imposed. Moreover,
there are also two chargeless superfields $W_1$ and $W_2$ with
specific Grassmann and harmonic shortness constraints describing
the same multiplet. All these six strength superfields are shown
to appear naturally in the solution of $\cN{=}4$ SYM constraints
with the help of $USp(4)/(U(1)\times U(1))$ harmonic variables.
Apart from the $\cN{=}4$ SYM multiplet, the superparticle leads to
the superfield realizations of $\cN{=}4$ gravitino multiplet (with
highest helicity 3/2) and $\cN{=}4$ supergravity multiplet (with
highest helicity 2). These multiplets are represented by
chiral-analytic and $\cN{=}4$ chiral superfields in harmonic
superspace, respectively, with appropriate Grassmann and harmonic
constraints.

The quantization of the $\cN{=}4$ harmonic superparticle shows
some new possibilities for studying the $\cN{=}4$ SYM theory
directly in $\cN{=}4$ harmonic superspace with $USp(4)/(U(1)\times
U(1))$ harmonics. The $USp(4)$ group is very suitable for this
purpose, particularly because of its invariant antisymmetric
2-tensor, which raises and lowers the R-symmetry indices,
similarly as the $\varepsilon$-tensor in the $SU(2)$ group. The
corresponding $USp(4)$ harmonic superspace possesses a specific
conjugation generalizing the usual complex conjugation, and
harmonic projections of $\cN{=}4$ superfield strengths appear real
under this conjugation. Therefore the $USp(4)$ harmonics seem to
be very useful for studying the $\cN{=}4$ SYM model in harmonic
superspace. Moreover, as is sketched in the last subsection, it is
very straightforward to build some gauge invariant actions in
$\cN{=}4$ harmonic superspace with $USp(4)$ harmonics. We propose
actions which contain
$F^4$ term in the bosonic sector. These actions are written in an
analytic superspace with an integration over half of the Grassmann
variables of $\cN{=}4$ superspace, such that the analytic measure
is dimensionless.
The $F^4$ term may be interpreted as the quartic term in an
$\cN{=}4$ Born-Infeld action. It is interesting to note that it
allows for a very simple scale-invariant generalization which
corresponds to the leading term in the low-energy effective action
of the $\cN{=}4$ SYM model. However, these issues require deeper
investigations.

To conclude, we have constructed the $\cN{=}4$ massive and
massless superparticle models and developed their quantization. As
a result we found superfields in $\cN{=}4$ harmonic superspace
with $USp(4)/(U(1)\times U(1))$ harmonics which describe the basic
on-shell $\cN{=}4$ multiplets. These superfields allow one to
write some $\cN{=}4$ superfield actions corresponding to $\cN{=}4$
SYM theory.

It would be very interesting to develop systematically the
ideas presented in Appendix 3, where we show that the analytic
potentials, which serve as gauge connections for the harmonic
derivatives, can be used for constructing some superfield actions.

\vspace{5mm} {\bf Acknowledgements.} The authors are grateful to
E.A. Ivanov, N.G. Pletnev and D.P. Sorokin for stimulating
discussions. The present work is supported particularly by INTAS
grant, project No 05-1000008-7928, by RFBR grants, projects No
06-02-16346 and No 08-02-90490, by DFG grant, project No 436
RUS/113/669/0-3 and grant for LRSS, project No 2553.2008.2. I.B.S.
acknowledges the support from INTAS grant, project No
06-1000016-6108.

\appendix
\def\theequation{A.\arabic{equation}}
\def\thesection{}
\def\thesubsection{\arabic{subsection}}
\setcounter{equation}{0}
\section{Appendices}
\subsection{Commutation relations in $usp(4)$ algebra}
Let us introduce the operators
\be
T^{ij}=\Omega^{k(i}u^{j)}{}_l\frac\partial{\partial u^k{}_l},
\label{aa1}
\ee
where $u^i{}_j$ are the harmonic variables (\ref{e8}).
It well known, that the operators (\ref{aa1}) obey the commutation
relations of $usp(4)$ algebra (see, e.g., \cite{FerSok}),
\be
[T^{ij},T^{kl}]=\Omega^{i(k}T^{l)j}+\Omega^{j(k}T^{l)i}.
\label{aa2}
\ee
In the present work we prefer to use
the operators (\ref{e27}) which differ from (\ref{aa1}) only by
constants and therefore also span the $usp(4)$ algebra.
Their commutation relations are given by
\begin{align}
&[D^{(++,0)},D^{(--,0)}]=S_1,&&
  [D^{(0,++)},D^{(0,--)}]=S_2,\nn\\
&[D^{(+,+)},D^{(-,-)}]=S_1+S_2,&&
  [D^{(+,-)},D^{(-,+)}]=S_2-S_1,\nn\\
&[D^{(+,-)},D^{(+,+)}]=2D^{(++,0)},&&
  [D^{(+,-)},D^{(-,-)}]=-2D^{(0,--)},\nn\\
&[D^{(+,-)},D^{(++,0)}]=0,&&
  [D^{(+,-)},D^{(--,0)}]=-D^{(-,-)},\nn\\
&[D^{(+,-)},D^{(0,++)}]=D^{(+,+)},&&
  [D^{(-,-)},D^{(--,0)}]=0, \nn\\
&[D^{(+,+)},D^{(++,0)}]=0,&&
  [D^{(+,+)},D^{(--,0)}]=-D^{(-,+)},\nn\\
&[D^{(+,+)},D^{(0,++)}]=0,&&
  [D^{(+,+)},D^{(0,--)}]=D^{(+,-)},\nn\\
&[D^{(++,0)},D^{(0,++)}]=0,&&
  [D^{(++,0)},D^{(0,--)}]=0,\nn\\
&[D^{(+,+)},D^{(-,+)}]=2D^{(0,++)},&&
  [D^{(++,0)},D^{(-,+)}]=D^{(+,+)},\nn\\
&[D^{(++,0)},D^{(-,-)}]=D^{(+,-)},&&
  [D^{(0,++)},D^{(-,+)}]=0,\nn\\
&[D^{(0,++)},D^{(-,-)}]=-D^{(-,+)},&&
  [D^{(0,++)},D^{(--,0)}]=0\nn\\
&[D^{(-,+)},D^{(0,--)}]=D^{(-,-)},&&
  [D^{(-,+)},D^{(-,-)}]=2D^{(--,0)}\nn\\
&[D^{(+,-)},D^{(0,--)}]=0,&&
  [D^{(-,+)},D^{(--,0)}]=0,\nn\\
&[D^{(-,-)},D^{(0,--)}]=0,&&
  [D^{(--,0)},D^{(0,--)}]=0,\nn\\
&[S_1,D^{(++,0)}]=2D^{(++,0)},&&
 [S_2,D^{(++,0)}]=0,\nn\\
&[S_1,D^{(0,++)}]=0,&&
 [S_2,D^{(0,++)}]=2D^{(0,++)}\nn\\
&[S_1,D^{(--,0)}]=-2D^{(--,0)},&&
 [S_2,D^{(--,0)}]=0,\nn\\
&[S_1,D^{(0,--)}]=0,&&
 [S_2,D^{(0,--)}]=-2D^{(0,--)}\nn\\
&[S_1,D^{(+,+)}]=D^{(+,+)},&&
 [S_2,D^{(+,+)}]=D^{(+,+)},\nn\\
&[S_1,D^{(-,-)}]=-D^{(-,-)},&&
 [S_2,D^{(-,-)}]=-D^{(-,-)},\nn\\
&[S_1,D^{(+,-)}]=D^{(+,-)},&&
 [S_2,D^{(+,-)}]=-D^{(+,-)},\nn\\
&[S_1,D^{(-,+)}]=-D^{(-,+)},&&
 [S_2,D^{(-,+)}]=D^{(-,+)},\nn\\
&[S_1,S_2]=0.
\label{e22}
\end{align}

\subsection{Comment on the $D^\alpha D_\alpha$ constraint and $\kappa$-symmetries}
Here we explain the origin of the quadratic spinor constraints
(\ref{e102.4},\ref{e107},\ref{e115.1},\ref{e126}) appearing in the
massless case. All the considerations are valid for arbitrary
$\cN$, however only $\cN{=}4$ superparticle was studied in this
work.

Consider the Lagrangian of massless superparticle (\ref{e100})
with the constraints (\ref{e53},\ref{e57}), which in the massless
case read
\be
p^2\approx 0,\label{a2.1}
\ee
\be
D^i_\alpha=-\pi^i_\alpha+ip_m(\sigma^m\bar\theta^i)_\alpha\approx0,\qquad
\bar
D_{i\dot\alpha}=\bar\pi_{i\dot\alpha}-ip_m(\theta_i\sigma^m)_{\dot\alpha}\approx
0.
\label{a2.2}
\ee
The transformations of $\kappa$-symmetry (\ref{kappa}) in the
massless case
\bea
\delta_\kappa\theta_{i\alpha}&=&-ip_m\sigma^m_{\alpha\dot\alpha}
 \bar\kappa^{\dot\alpha}_i,
\qquad
\delta_\kappa\bar\theta^i_{\dot\alpha}=ip_m\kappa^{i\alpha}\sigma^m_{\alpha\dot\alpha},
\nn\\
\delta_\kappa x^m&=&i\delta_\kappa\theta_i\sigma^m\bar\theta^i
-i\theta_i\sigma^m\delta_\kappa\bar\theta^i,\qquad
\delta_\kappa e=-4(\bar\kappa_{i\dot\alpha}\dot{\bar\theta}{}^{i\dot\alpha}
+\dot\theta^\alpha_i\kappa^i_\alpha).
\label{a-kappa}
\eea
are generated by
\be
\delta_\kappa=\kappa^{i\alpha} [\psi_{i\alpha},.\}_P
-\bar\kappa^{\dot\alpha}_i[ \bar\psi^i_{\dot\alpha},.\}_P,
\label{a-var}
\ee
where $[.,.\}_P$ is the graded Poisson bracket and
\be
\psi_{i\alpha}=ip_m\sigma^m_{\alpha\dot\alpha}\bar
D^{\dot\alpha}_i\approx 0,\qquad
\bar\psi^i_{\dot\alpha}=-ip_m\sigma^m_{\alpha\dot\alpha}D^{i\alpha}\approx0
\label{a-psi}
\ee
are the generators of $\kappa$-symmetry which are first-class
constraints.

We point out that the transformations (\ref{a-kappa}) leave
the superparticle Lagrangian invariant ($\delta_\kappa L_1=0$) for arbitrary local
parameters $\kappa_\alpha$, $\bar\kappa_{\dot\alpha}$. In particular, we
can take
\be
\kappa^i_\alpha=k^{ij}\theta_{j\alpha},\qquad
\bar\kappa_{i\dot\alpha}=\bar k_{ij}\bar \theta^j_{\dot\alpha},
\label{a-choose}
\ee
where $k^{ij}=k^{ji}$, $\bar k_{ij}=\bar k_{ji}$ are new local bosonic parameters.
The transformations (\ref{a-kappa}) read now
\bea
\delta_k\theta_{i\alpha}&=&-i\bar k_{ij} p_m\sigma^m_{\alpha\dot\alpha}
\bar\theta^{j\dot\alpha},\qquad
\delta_k\bar\theta^i_{\dot\alpha}=ik^{ij}p_m\theta_j^\alpha\sigma^m_{\alpha\dot\alpha},\nn\\
\delta_k x^m&=&-p^m[k^{ij}\theta^\alpha_i\theta_{j\alpha}
 +\bar
 k_{ij}\bar\theta^i_{\dot\alpha}\bar\theta^{j\dot\alpha}],\qquad
\delta_k e=-4(k^{ij}\dot\theta{}_i^\alpha\theta_{j\alpha}+\bar
k_{ij}\bar\theta^i_{\dot\alpha}\dot{\bar\theta}{}^{j\dot\alpha}).
\label{a-z}
\eea
They are generated by the variation
\be
\delta_k=k^{ij}[K_{ij},.]_P+\bar k_{ij}[\bar K^{ij},.]_P,
\label{a-var1}
\ee
where $K_{ij}$, $\bar K^{ij}$ are new first-class constraints,
\be
K_{ij}=-ip_m\theta_{(i}^\alpha\sigma^m_{\alpha\dot\alpha}\bar\pi_{j)}^{\dot\alpha}
\approx0,\qquad
\bar
K^{ij}=-ip_m\pi^{(i\alpha}\sigma^m_{\alpha\dot\alpha}\bar\theta^{j)\dot\alpha}
\approx0.
\label{a-Z}
\ee
Of course, these constraints are not independent, but follow from
(\ref{a-psi}) upon contractions with Grassmann variables.
With the use of (\ref{a2.1},\ref{a2.2}) the constraints (\ref{a-Z}) can be
rewritten as
\be
\bar\pi_{i\dot\alpha}\bar\pi_j^{\dot\alpha}\approx0,\qquad
\pi^{i\alpha}\pi^j_\alpha\approx0,
\label{a-Z1}
\ee
or as
\be
\bar D_{ij}\equiv\bar D_{i\dot\alpha}\bar D_j^{\dot\alpha}\approx0,\qquad
D^{ij}\equiv D^{i\alpha} D^j_\alpha\approx0.
\label{a-Z2}
\ee
Therefore the constraints (\ref{a-Z2}) can also be imposed on the states upon
quantization.

Note also that the generators of $\kappa$-symmetries appear from
the commutators of the constraints (\ref{a2.2}) and (\ref{a-Z}),
\be
[K_{ij},D^k_\alpha]_P=\psi_{(i\alpha}\delta^k_{j)},\qquad
[\bar K^{ij},\bar D_{k\dot\alpha}]_P=\bar\psi^{(i}_{\dot\alpha}
\delta^{j)}_k.
\label{a-com}
\ee
\be
[\bar D_{ij},D^k_\alpha]_P=4\psi_{(i\alpha}\delta^k_{j)},\qquad
[D^{ij},\bar D_{k\dot\alpha}]_P=4\bar\psi^{(i}_{\dot\alpha}
\delta^{j)}_k.
\label{a-com1}
\ee
Therefore one can consider only the constraints (\ref{a2.2}) and (\ref{a-Z2})
since the $\kappa$-symmetry ones (\ref{a-kappa}) follow from
their algebra.

As a result, there is a bosonic version of $\kappa$-symmetry if
we choose the parameters of $\kappa$-transformations being
proportional to Grassmann coordinates. However, not all the parameters
$k_{ij}$, $\bar k^{ij}$ are independent. Indeed, a symmetric
$\cN\times\cN$ matrix has $\cN(\cN+1)/2$ independent elements
that is too much in comparison with the number of first-class
superparticle
constraints. However, there is $U(\cN)$ R-symmetry which rotates
indices $i,j$ of all objects. It is well-known, that any symmetric
complex matrix $k^{ij}$ can be brought to the diagonal form by $U(\cN)$
rotations \cite{Zumino},
\be
u k u^{\rm T}={\rm diag}(d_1,\ldots, d_{\cN}),\qquad u\in U(\cN).
\label{a-rot}
\ee
After such a rotation, there are only $\cN$
independent real parameters in the matrix $k$.

We point out the importance of accounting the
constraints (\ref{a-Z2}) in the Gupta-Bleuler quantization of a
superparticle despite they are not
independent but follow from spinorial constraints (\ref{a2.2}) and
$\kappa$-symmetries (\ref{a-psi}).
Recall that the spinorial
constraints (\ref{a2.2}) consist of $2\cN$ first-class and $2\cN$
second-class constraints (which can not be separated explicitly)
while the $\kappa$-symmetry constraints (\ref{a-psi}) have
effectively $2\cN$ first-class constraints (since they are infinitely
reducible). Indeed, on the example of a chiral superfield
(\ref{e106}) it is easy to see that the constraints (\ref{e102.3})
are not sufficient and the constraint (\ref{e107}) must be imposed
to achieve the correct component structure (\ref{e108}). This is a
feature of Gupta-Bleuler quantization approach in which we take
into account only the constraint $\bar D_{i\dot\alpha}\approx0$
corresponding to the
chiral superfield but not $D^i_\alpha\approx0$. As a result, some
of the first class constraints originating from
$D^i_\alpha\approx0$ remain unaccounted until one has not imposed
the constraints (\ref{e107}).

Finally, we note that the analogous quadratic spinorial
constraints (\ref{e81}) also originate from the $\kappa$-symmetry
if one passes to the bosonic parameters as in (\ref{a-choose}).

\subsection{$F^2$ term in $\cN{=}4$ $USp(4)/(U(1)\times U(1))$
harmonic superspace} It is well known that the classical actions
in $\cN{=}2$ and $\cN{=}3$ SYM models can be written in terms of
unconstrained superfields (prepotentials) in harmonic superspace
\cite{HSS,Harm1}. A similar superfield formulation for $\cN{=}4$
SYM model is very desirable since it may be fruitful for studying
quantum aspects of this model. One can hope that the $\cN{=}4$
harmonic superspace with $USp(4)/(U(1)\times U(1))$ harmonic
variables may be useful for this purpose. In particular, here we
propose some functional built of analytic prepotentials which
gives $F^2$ term in components.

In principle, one can try to construct some actions in different
subspaces of full $\cN{=}4$ superspace. However, the superspaces
with eight Grassmann variables are more promising since the
corresponding integration measure is dimensionless. Indeed, if one
tries to construct an action in full $\cN{=}4$ superspace one
needs the dimensionful constant since the corresponding measure is
of dimension $+4$. Therefore we consider the analytic subspaces of
$\cN{=}4$ superspace which are singled out by the covariant spinor
derivatives in different lines in (\ref{l14}). Our aim now is to
develop the differential geometry in one of such analytic
subspaces and to build possible gauge invariant superfield
functionals.

Consider the analytic subspace in $\cN{=}4$ harmonic superspace
with coordinates\\ $\{x^m_A,\theta^{(+,0)}_\alpha,
\theta^{(0,+)}_\alpha, \bar\theta^{(+,0)}_{\dot\alpha},
\bar\theta^{(0,+)}_{\dot\alpha}, u\}$, where $x^m_A$ is given by
(\ref{e86}). As is shown in (\ref{e115}), the Grassmann
derivatives \be D^{(+,0)}_\alpha,\quad D^{(0,+)}_\alpha,\quad \bar
D^{(+,0)}_{\dot\alpha},\quad \bar D^{(0,+)}_{\dot\alpha}
\label{ap0} \ee single out the analytic superfields. The set of
derivatives (\ref{ap0}) is invariant under the commutators with
the following harmonic derivatives \be D^{(++,0)},\quad
D^{(0,++)},\quad D^{(+,+)},\quad D^{(-,+)},\quad D^{(+,-)}.
\label{ap0.1} \ee Therefore we can introduce five analytic gauge
connections for the harmonic derivatives (\ref{ap0.1}), \bea
D^{(++,0)}&\to&\nabla^{(++,0)}=D^{(++,0)}+V^{(++,0)},\nn\\
D^{(0,++)}&\to&\nabla^{(0,++)}=D^{(0,++)}+V^{(0,++)},\nn\\
D^{(+,+)}&\to&\nabla^{(+,+)}=D^{(+,+)}+V^{(+,+)},\nn\\
D^{(+,-)}&\to&\nabla^{(+,-)}=D^{(+,-)}+V^{(+,-)},\nn\\
D^{(-,+)}&\to&\nabla^{(-,+)}=D^{(-,+)}+V^{(-,+)}.
\label{e144}
\eea
Clearly, these prepotentials contain too much component fields even
in the physical sector
and we have to introduce some constraints. One of the possible
ways to impose the off-shell constraints is to vanish the
prepotentials in the last two lines in (\ref{e144}),
\be
V^{(+,-)}=0,\qquad
V^{(-,+)}=0.
\label{e144_}
\ee
We point out that it is the constraints (\ref{e144_}) which do not lead to the equations
of motion for the component fields, but just reduce the number of
independent components. Since the derivatives
$D^{(+,-)}$, $D^{(-,+)}$ form $su(2)$ subalgebra in (\ref{e22}),
the constraints (\ref{e144_}) can be naturally resolved if one
uses the harmonic variables on $USp(4)/(SU(2)\times U(1))$ coset
rather than $USp(4)/(U(1)\times U(1))$. However, for our
considerations it is sufficient to work in the $USp(4)/(U(1)\times
U(1))$ harmonic superspace with the constrained prepotentials
(\ref{e144}).

The covariant derivatives (\ref{e144}) should satisfy the same
algebra (\ref{e22}) even with imposed constraints (\ref{e144_}).
It leads to the following constraints for the
prepotentials,
\bea
&&D^{(+,-)}V^{(++,0)}=0,\qquad
D^{(-,+)}V^{(0,++)}=0,
\nn\label{e144.1}\\
&&
D^{(+,-)}D^{(+,-)}V^{(+,+)}=0,\qquad
D^{(-,+)}D^{(-,+)}V^{(+,+)}=0,\label{e144.2}\nn\\
&&(D^{(-,+)}D^{(+,-)}+2)V^{(+,+)}=0.
\label{e144.3}
\eea
Moreover, the prepotentials $V^{(++,0)}$, $V^{(0,++)}$,
$V^{(+,+)}$ can be expressed through each other and only
$V^{(+,+)}$ is independent,
\bea
&&D^{(-,+)}V^{(++,0)}=-V^{(+,+)},\qquad
D^{(+,-)}V^{(+,+)}=2V^{(++,0)},
\label{e144.4}\nn\\
&&D^{(+,-)}V^{(0,++)}=V^{(+,+)},\qquad
D^{(-,+)}V^{(+,+)}=-2V^{(0,++)}.
\label{e144.5}
\eea
There are also the following reality properties for the
prepotentials,
\be
\widetilde{V^{(+,+)}}=V^{(+,+)},\quad
\widetilde{V^{(++,0)}}=V^{(0,++)},\quad
\widetilde{V^{(0,++)}}=V^{(++,0)}.
\label{e144.6}
\ee

The analytic prepotentials define the harmonic field strengths,
\bea
F^{(2,2)}&=&[\nabla^{(++,0)},\nabla^{(0,++)}]=
D^{(++,0)}V^{(0,++)}-D^{(0,++)}V^{(++,0)}+[V^{(++,0)},V^{(0,++)}],\nn\\
F^{(3,1)}&=&[\nabla^{(++,0)},\nabla^{(+,+)}]=
D^{(++,0)}V^{(+,+)}-D^{(+,+)}V^{(++,0)}+[V^{(0,++)},V^{(+,+)}],\nn\\
F^{(1,3)}&=&[\nabla^{(+,+)},\nabla^{(0,++)}]=
D^{(+,+)}V^{(0,++)}-D^{(0,++)}V^{(+,+)}+[V^{(+,+)},V^{(0,++)}],
\label{e145}
\eea
which are analytic and gauge covariant (or invariant in
the Abelian case),
\be
\delta F^{(2,2)}=[\lambda, F^{(2,2)}],\quad
\delta F^{(3,1)}=[\lambda, F^{(3,1)}],\quad
\delta F^{(1,3)}=[\lambda, F^{(1,3)}]
\label{e147}
\ee
under the following gauge transformations of the prepotentials
\bea
\delta V^{(++,0)}=-\nabla^{(++,0)}\lambda,\quad
\delta V^{(0,++)}=-\nabla^{(0,++)}\lambda,\quad
\delta V^{(+,+)}=-\nabla^{(+,+)}\lambda.
\label{e146}
\eea
Here $\lambda$ is a real analytic gauge parameter constrained by
\be
D^{(+,-)}\lambda=D^{(-,+)}\lambda=0.
\label{e146.1}
\ee
Owing to the constraints (\ref{e144.3}) these
strengths are related to each other,
\be
D^{(+,-)}F^{(2,2)}=F^{(3,1)},\quad
D^{(-,+)}F^{(2,2)}=F^{(1,3)},\quad
D^{(+,-)}F^{(1,3)}=D^{(-,+)}F^{(3,1)}=-2F^{(2,2)}.
\label{e147.1}
\ee

We stress that the constraints (\ref{e146.1}) are purely
algebraical and mean that $\lambda$ depends on $USp(4)/(SU(2)\times
U(1))$ harmonics. One can avoid all these constraints by using
the superfields in $USp(4)/(SU(2)\times U(1))$ harmonic
superspace.

Now we apply the strength superfields (\ref{e145}) to build
a gauge invariant action in analytic subspace,
\be
S_2=-\,\tr\int d\zeta^{(-4,-4)}F^{(2,2)}F^{(2,2)}
=-\frac12\tr\int d\zeta^{(-4,-4)}F^{(3,1)}F^{(1,3)},
\label{e149}
\ee
where $d\zeta^{(-4,-4)}$ is the analytic measure,
\be
d\zeta^{(-4,-4)}=\frac1{2^8}d^4x_A du(D^{(-,0)})^2 (D^{(0,-)})^2 (\bar D^{(-,0)})^2
(\bar D^{(0,-)})^2.
\label{e143}
\ee
The integration over harmonic variables is defined by the
following rules
\bea
&&\int du\, 1=1,\nn\\
&&\int du\, f^{(s_1,s_2)} =0, \quad\mbox{if}\quad s_1\ne0,\
s_2\ne0,\nn\\
&&\int du\,(\mbox{irreducible harmonic tensor})=0,
\label{e142}
\eea
where $f^{(s_1,s_2)}$ is some function of $USp(4)/(U(1)\times
U(1))$ harmonic variables. The rigorous grounds for such rules of
harmonic integrals are given in the book \cite{Book} for the case
of $SU(2)/U(1)$ harmonic variables and in \cite{Harm1,Delduc} for
$SU(3)/(U(1)\times U(1))$ harmonics. Here we just generalize these
constructions for the $USp(4)/(U(1)\times U(1))$ coset.

The action (\ref{e149}) is supersymmetric and gauge invariant by
construction and contain the component fields with the spins
(helicity) not higher than one. In particular, it is easy to find
the vector field in its component structure by considering the
following term in the prepotential $V^{(+,+)}$,
\be
V^{(+,+)}=\frac i2[\theta^{(+,0)}\sigma^m\bar\theta^{(0,+)}
 +\theta^{(0,+)}\sigma^m\bar\theta^{(+,0)}]A_m+\ldots.
\label{ap3_} \ee The real vector field $A_m$ turns into the
Maxwell strength in the $F^{(2,2)}$ superfield (in the Abelian
case), \be F^{(2,2)}=(\theta^{(+,0)}\sigma^m\bar\theta^{(+,0)})
(\theta^{(0,+)}\sigma^n\bar\theta^{(0,+)})F_{mn} +\ldots
\label{ap3} \ee and leads to the Maxwell term in the action
(\ref{e149}), \be S_2=-\frac14\int d^4x \,  F^{mn}F_{mn}+\ldots.
\label{ap4} \ee The dots stand here for the other component terms
in the action. As a result we see that the action contains
$\cN{=}4$ vector multiplet in component decomposition and may have
some relation to the $\cN{=}4$ supergauge theory. However, one can
check that the action (\ref{e149}) contains much more propagating
degrees of freedom than a single $\cN{=}4$ SYM multiplet.
Therefore to make the action (\ref{e149}) physical one needs more
superfield constraints for the prepotentials and this issue
requires further studies.

Another way in the seek of unconstrained $\cN{=}4$ SYM action in
the $USp(4)$ harmonic superspace is the use of other analytic
subspaces in $\cN{=}4$ superspace. Of particular interest may by
the analytic subspace with coordinates
$\{x_A^m,\theta^{(+,0)}_\alpha,\theta^{(-,0)}_\alpha,
\bar\theta^{(0,+)}_{\dot\alpha},\bar\theta^{(0,-)}_{\dot\alpha},u\}$,
where $x^m_A$ is given by (\ref{x6}). The corresponding analytic
measure is chargeless, \be d\zeta=\frac1{2^8}d^4x_A du
(D^{(+,0)})^2 (D^{(-,0)})^2 (\bar D^{(0,+)})^2 (\bar D^{0,-})^2.
\label{e152a} \ee One can use other prepotentials and field
strengths in this subspace for constructing the invariant actions.
We leave these questions for further studies.


\begin{thebibliography}{99}
\addtolength{\itemsep}{-5pt}
\bibitem{AdsCFT}
O. Aharony, S.S. Gubser, J.M. Maldacena, H. Ooguri, Y. Oz,
{\it Large N field theories, string theory and gravity},
Phys. Rept. 323 (2000) 183, {\tt hep-th/9905111}.
%
\bibitem{N1} S.J. Gates, M.T. Grisaru, M. Ro\v cek, W. Siegel,
{\it Superspace: Or one thousand and one lessons in
supersymmetry}, Benjamin/Cummings, 1983, 548 p.
%
\bibitem{HSS}
A. Galperin, E. Ivanov, V. Ogievetsky, E. Sokatchev, {\it Harmonic
superspace: Key to N=2 supersymmetric theories}, JETP Lett.
40 (1984) 912;\\
A. Galperin, E. Ivanov, S. Kalitzin, V. Ogievetsky, E. Sokatchev,
{\it Unconstrained N=2 matter, Yang-Mills and supergravity theories in
harmonic superspace}, Class. Quant. Grav. 1 (1984) 469.
%
\bibitem{HSS1}
A. Galperin, E.A. Ivanov, V. Ogievetsky, E. Sokatchev, {\it
Harmonic supergraphs. Green functions}, Class. Quant. Grav. 2 (1985)
601; \\
{\it Harmonic supergraphs. Feynman rules and examples},
Class. Quant. Grav. 2 (1985) 617.
%
\bibitem{Harm1} A. Galperin, E. Ivanov, S. Kalitzin, V.
Ogievetsky, E.~Sokatchev, {\it N=3 Supersymmetric gauge theory},
Phys. Lett. B151 (1985) 215;\\
{\it Unconstrained off-shell N=3
supersymmetric Yang-Mills theory}, Class. Quant. Grav. 2
(1985) 155.
%
\bibitem{Book} A. Galperin, E. Ivanov, V. Ogievetsky, E.
Sokatchev, {\it Harmonic Superspace}, UK: Cambridge Univ. Press, 2001, 306 p.
%
\bibitem{N4:(} E. Ahmed, S. Bedding, C.T. Card, M. Dumbrell,
M. Nouri-Moghadam, J.G. Taylor, {\it On N=4 supersymmetric
Yang-Mills in harmonic superspace}, J. Phys. A18 (1985) 2095.
%
\bibitem{Bandos1} I.A. Bandos, {\it Solution of linear equations in spaces of harmonic
variables}, Theor. Math. Phys. 76 (1988) 783
[Teor. Mat. Fiz. 76 (1988) 169].
%
\bibitem{Howe95}
G.G. Hartwell, P.S. Howe, {\it (N, p, q) harmonic superspace},
Int. J. Mod. Phys. A10 (1995) 3901, {\tt hep-th/9412147};\\
P. Heslop, P.S. Howe, {\it On harmonic superspaces and superconformal
fields in four dimensions}, Class. Quant. Grav. 17 (2000) 3743,
{\tt hep-th/0005135}.
%
\bibitem{Ferrara} L. Andrianopoli, S. Ferrara, E. Sokatchev, B. Zupnik,
{\it Shortening of primary operators in N extended SCFT(4) and
harmonic superspace analyticity}, Adv. Theor. Math. Phys. 3 (1999) 1149,
{\tt hep-th/9912007}.
%
\bibitem{Fer99} S. Ferrara, E. Sokatchev, {\it Short representations of
SU(2,2/N) and harmonic superspace analyticity}, Lett. Math. Phys.
52 (2000) 247, {\tt hep-th/9912168}.
%
\bibitem{DHHK} J.M. Drummond, P.J. Heslop, P.S. Howe, S.F.
Kerstan, {\it Integral invariants in N=4 SYM and the effective action
for coincident D-branes}, JHEP 0308 (2003) 016, {\tt
hep-th/0305202}.
%
\bibitem{superembedding} D.P. Sorokin, {\it Superbranes and
superembeddings}, Phys. Rept. 329 (2000) 1, {\tt hep-th/9906142}.
%
\bibitem{Casalbuoni}
R. Casalbuoni, {\it The classical mechanics for Bose-Fermi systems},
Nuovo Cim. A33 (1976) 389.
%
\bibitem{VolkovPashnev}
A.I. Pashnev and D.V. Volkov, {\it Supersymmetric Lagrangian
for particles in proper time}, Theor. Math. Phys. 44 (1980) 770
[Teor. Mat. Fiz. 44 (1980) 321].
%
\bibitem{BS} L. Brink, J.H. Schwarz, {\it Quantum superspace},
Phys. Lett. B100 (1981) 310.
%
\bibitem{Luk1} J.A. de Azc\'arraga, J. Lukierski,
{\it Supersymmetric particles with internal symmetries and central
charges}, Phys. Lett. B113 (1982) 170;\\
A. Frydryszak, J. Lukierski, {\it N=2 massive matter multiplet from
quantization of extended classical mechanics}, Phys. Lett.
B117 (1982) 51;\\
 J.A. de Azc\'arraga, J. Lukierski, {\it
Supersymmetric particles in N=2 superspace: Phase-space variables
and Hamiltonian dynamics}, Phys. Rev. D28 (1983) 1337.
%
\bibitem{Sorokin1} V.P. Akulov, D.P. Sorokin, I.A. Bandos,
{\it Particle mechanics in harmonic superspace},
Mod. Phys. Lett. A3 (1988) 1633.
%
\bibitem{Sorokin2} V.P. Akulov, I.A. Bandos, D.P. Sorokin,
{\it Particle in harmonic N=2 superspace},
Sov. J. Nucl. Phys. 47 (1988) 724 [Yad. Fiz. 47 (1988)
1136-1146].
%
\bibitem{stv}
D.P. Sorokin, V.I. Tkach, D.V. Volkov, {\it Superparticles,
twistors and Siegel symmetry}, Mod. Phys. Lett.  A4 (1989) 901.
%
\bibitem{Lus}
L. Lusanna, B. Milewski, {\it N=2 Super Yang-Mills and supergravity
constraints from coupling to a supersymmetric particle},
Nucl. Phys. B247 (1984) 396;\\
J.A. Shapiro, C.C. Taylor, {\it Superspace supergravity from the
superstring}, Phys. Lett. B186 (1987) 69.
%
\bibitem{BS1} I.L. Buchbinder, I.B. Samsonov, {\it N=3 Superparticle
model}, {\tt arXiv:0801.4907 [hep-th]}.
%
\bibitem{Fayet79} P. Fayet, {\it Spontaneous generation of massive multiplets and
central charges in extended supersymmetric theories}, Nucl. Phys.
B149 (1979) 137.
%
\bibitem{FS} S. Ferrara, C.A. Savoy, B. Zumino, {\it
General massive multiplets in extended supersymmetry}, Phys. Lett. B100
(1981) 393.
%
\bibitem{IKNO} E. Ivanov, S. Kalitzin, N. Ai Viet, V. Ogievetsky,
{\it Harmonic superspaces and extended supersymmetry: The calculus
of harmonic variables}, J. Phys. A18 (1985) 3433.
%
\bibitem{Sokatchev96} E. Sokatchev, {\it An action for N=4 supersymmetric
selfdual Yang-Mills theory}, Phys. Rev. D53 (1996) 2062, {\tt
hep-th/9509099}.
%
\bibitem{FerSok} S. Ferrara, E. Sokatchev,
{\it Superconformal interpretation of BPS states in AdS
geometries}, Int. J. Theor. Phys. 40 (2001) 935, {\tt
hep-th/0005151};\\
{\it Conformal superfields and
BPS states in AdS(4/7) geometries}, Int. J. Mod. Phys. B14
(2000) 2315, {\tt hep-th/0007058};\\
{\it Representations of (1,0) and (2,0) superconformal algebras
in six-dimensions: Massless and short superfields},
Lett. Math. Phys. 51 (2000) 55, {\tt hep-th/0001178};\\
{\it Universal properties of superconformal OPEs for 1/2 BPS
operators in 3$\leq$D$\leq$6}, New J. Phys. 4 (2002) 2, {\tt
hep-th/0110174}.
%
\bibitem{corr} P.J. Heslop, P.S. Howe, {\it A Note on composite operators in N=4
SYM}, Phys. Lett. B516 (2001) 367, {\tt hep-th/0106238};\\
{\it OPEs and three-point correlators of protected operators in N=4
SYM}, Nucl. Phys. B626 (2002) 265, {\tt hep-th/0107212};\\
B. Eden, E. Sokatchev, {\it On the OPE of 1/2 BPS short operators in N=4
SCFT(4)}, Nucl. Phys. B618 (2001) 259, {\tt hep-th/0106249};\\
G. Arutyunov, F.A. Dolan, H. Osborn, E. Sokatchev, {\it
Correlation functions and massive Kaluza-Klein modes in the AdS/CFT
correspondence}, Nucl. Phys. B665 (2003) 273, {\tt
hep-th/0212116};\\
E. D'Hoker, P. Heslop, P. Howe, A.V. Ryzhov, {\it
Systematics of quarter BPS operators in N=4 SYM}, JHEP 0304 (2003) 038,
{\tt  hep-th/0301104}.
%
\bibitem{Zumino} B. Zumino, {\it Normal forms of complex
matrices}, J. Math. Phys. 3 (1962) 1055.
%
\bibitem{Sokathev86} E. Sokatchev, {\it Light cone harmonic
superspace and its applications}, Phys. Lett. B169 (1986) 209;\\
{\it Harmonic superparticle}, Class. Quant. Grav. 4 (1987) 237;\\
F. Delduc, A. Galperin, E. Sokatchev, {\it Lorentz harmonic (super)fields
 and (super)particles}, Nucl. Phys. B368 (1992) 143.
%
\bibitem{Bandos2} I.A. Bandos, {\it Superparticle in Lorentz harmonic
superspace}, Sov. J. Nucl. Phys. 51 (1990) 906 [Yad. Fiz. 51 (1990) 1429];\\
{\it Spinor moving frame, M0-brane covariant BRST quantization and
intrinsic complexity of the pure spinor approach}, Phys. Lett. B659
(2008) 388, {\tt arXiv:0707.2336 [hep-th]};\\
{\it D=11 massless superparticle covariant quantization, pure spinor
BRST charge and hidden symmetries}, Nucl. Phys. B796 (2008) 360,
{\tt arXiv:0710.4342 [hep-th]}.
%
\bibitem{N2constr} R. Grimm, M. Sohnius, J. Wess, {\it Extended supersymmetry
and gauge theories}, Nucl. Phys. B133 (1978) 275;\\
M.F. Sohnius, {\it Bianchi identities for supersymmetric gauge
theories}, Nucl. Phys. B136 (1978) 461;\\
M.F. Sohnius, {\it Supersymmetry and central charges},
Nucl. Phys. B138 (1978) 109;\\
P. Howe, K.S. Stelle, P.K. Townsend, {\it
Supercurrents}, Nucl. Phys. B192 (1981) 332;\\
{\it The relaxed hypermultiplet:
An unconstrained N=2 superfield theory}, Nucl. Phys. B214 (1983)
519.
%
\bibitem{Echaya}
I.L. Buchbinder, S.M. Kuzenko, B.A. Ovrut,
{\it On the D=4, N=2 nonrenormalization theorem},
Phys. Lett. B433 (1998) 335, {\tt hep-th/9710142};\\
I.L. Buchbinder, E.I. Buchbinder, S.M. Kuzenko, {\it Non-holomorphic effective potential
in N=4 SU(n) SYM}, Phys. Lett. B446 (1999) 216, {\tt hep-th/9810239}; \\
E.I. Buchbinder, I.L. Buchbinder, E.A. Ivanov, S.M. Kuzenko, B.A.
Ovrut, {\it Low-energy effective action in N=2
supersymmetric field theories}, Phys. Part. Nucl. 32 (2001) 641;\\
I.L. Buchbinder, E.A. Ivanov, {\it Complete N=4 structure of
low-energy effective action in super yang-Mills theories},
Phys. Lett. B524 (2002) 208, {\tt hep-th/0111062};\\
I.L. Buchbinder, E.A. Ivanov, A.Yu. Petrov, {\it Complete low-energy
effective action in N=4 SYM: a direct N=2 supergraph calculation},
Nucl. Phys. B653 (2003) 64, {\tt hep-th/0210241};\\
I.L. Buchbinder, N.G. Pletnev, {\it Hypermultiplet dependence of
one-loop low-energy effective action in the N=2 superconformal
theories}, JHEP 0704 (2007) 096, {\tt hep-th/0611145}.
%
\bibitem{Zupnik07}
B.M. Zupnik, {\it Chern-Simons D=3, N=6 superfield theory}, Phys. Lett.
B660 (2008) 254, {\tt arXiv:0711.4680 [hep-th]};\\
{\it Chern-Simons theory in SO(5)/U(2) harmonic superspace},
{\tt arXiv:0802.0801 [hep-th]}.
%
\bibitem{Tseytlin} A.A. Tseytlin, {\it Born-Infeld action,
supersymmetry and string theory},
In *Shifman, M.A. (ed.): The many faces of the superworld*, p.
417-452, {\tt hep-th/9908105}.
%
\bibitem{IZ} E.A. Ivanov, B.M. Zupnik, {\it N=3 supersymmetric Born-Infeld
theory}, Nucl. Phys. B618 (2001) 3, {\tt hep-th/0110074}.
%
\bibitem{BISZ} I.L. Buchbinder, E.A. Ivanov, I.B. Samsonov, B.M.
Zupnik, {\it Scale invariant low-energy effective action in N=3 SYM
theory}, Nucl. Phys. B689 (2004) 91, {\tt hep-th/0403053}.
%
\bibitem{Delduc} F. Delduc, J. McCabe, {\it The quantization of N=3 super-Yang-Mills
off-shell in harmonic superspace}, Class. Quant. Grav. 6 (1989) 233.
%

\end{thebibliography}
\end{document}